\documentclass[%
 aip,pop,
 amsmath,amssymb,
 reprint,%
]{revtex4-2}

\usepackage[bookmarks=true,
            bookmarksopen=true,
            bookmarksnumbered=true,
            colorlinks=true,
            linkcolor=blue,
            citecolor=blue,
            urlcolor=blue]{hyperref}
\usepackage{tikz}
\usepackage{tikz-3dplot}
\usepackage{pgfplots}
\pgfplotsset{compat=1.18}
\usetikzlibrary{external, arrows.meta, decorations.markings, calc}

\usepackage{graphicx}%
\usepackage{dcolumn}%
\usepackage{bm}%
\usepackage{subcaption}

\usepackage[utf8]{inputenc}
\usepackage[T1]{fontenc}
\usepackage{newtxtext}
\usepackage{newtxmath}
\usepackage{etoolbox}
\usepackage{enumitem}
\usepackage{microtype}
\usepackage{siunitx}

\makeatletter
\def\@citess#1{\leavevmode\unskip\penalty\@M\ [\@cite{#1}{}]}
\makeatother

\makeatletter
\def\@email#1#2{%
 \endgroup
 \patchcmd{\titleblock@produce}
  {\frontmatter@RRAPformat}
  {\frontmatter@RRAPformat{\produce@RRAP{*#1\href{mailto:#2}{#2}}}\frontmatter@RRAPformat}
  {}{}
}%
\makeatother

\begin{document}

\preprint{AIP/123-QED}

\title[Ideal stability of the sheared-flow Z pinch]{On the ideal magnetohydrodynamic stability of the sheared-flow Z pinch}

\author{Daniel W. Crews}
\email{daniel.crews@zap.energy}
\affiliation{Zap Energy, Everett, WA 98203, USA}

\author{Jackson C. Turner}
\altaffiliation[Present address: ]{Harmoniqs, Brooklyn, NY, USA}
\affiliation{Zap Energy, Everett, WA 98203, USA}

\date{\today}

\begin{abstract}
Sheared-flow Z-pinch stability has previously been studied within ideal MHD primarily through growth rate calculations,
which find that even trans-Alfv\'{e}nic sheared flows apparently fail to suppress the kink instability.
Here, trans-Alfv\'{e}nic sheared flow is shown to stabilize the MHD kink but also to excite shear-driven instabilities
characteristic of high-Reynolds-number supersonic flow.
This distinction is evident from the dispersion relations underlying the growth rates,
computed as the analytic dispersion function in the complex-frequency plane.
Regularization splits this function into adiabatic and resonant parts describing how discrete modes
emerge from and interact with the continuous spectrum.
The Doppler-shifted flow continuum is then seen to interact with the interchange and kink instabilities in distinct ways.
For interchange, the continuum overlaps the instability branch at all wavenumbers,
so even sub-Alfv\'{e}nic sheared flow stabilizes profiles modestly beyond the interchange threshold.
The kink, by contrast, is shielded from the continuum by a frequency gap, and trans-Alfv\'{e}nic flow
is required to Doppler-shift the continuum into resonance with it, giving a geometric picture of the stabilization threshold.
But shear-driven instabilities arise at this same threshold, including reflection modes and an acoustic kink.
It is these shear-driven modes, not the original MHD instabilities,
that dominate the ideal-MHD spectrum in trans-Alfv\'{e}nic conditions.
The ideal analysis thus describes the stabilization mechanism while showing that the stability of the sheared-flow Z pinch
ultimately rests on non-ideal physics, such as finite orbit width and dissipation.
\end{abstract}

\maketitle

\section{Introduction}
\label{sec:introduction}

An ideal Z~pinch with enough flow to stabilize its MHD instabilities
has too much flow to be free of compressible ones.
This is because the MHD spectrum of the sheared-flow Z~pinch
contains both the familiar static plasma instabilities,
which sheared flow damps by Doppler shifting the wave continuum,
and compressible shear-driven instabilities excited through wave resonance by the same Doppler shifts.
These conclusions are reached by considering the linearized ideal MHD system as a boundary-value problem,
yielding an analytic dispersion function $\chi(\omega, k)$
of the complex frequency $\omega$,
whose structure in the complex frequency plane determines stability,
and which this work considers analytically and numerically.
A historical summary and paper organization follow.

\subsection{Historical summary}
Velocity shear modifies the spectrum of fluid instabilities
by coupling to convective structures through the Doppler shift,
a mechanism recognized since the classical theory of
Rayleigh and Kelvin~\cite{Drazin_2004} and exploited in tokamak
$\mathbf{E}\times\mathbf{B}$ shear suppression of convective transport.~\cite{Burrell_1997}

The Z-pinch MHD instabilities were identified in the 1950s.~\cite{Kruskal_1954}
For static plasmas, the energy principle established that the $m=0$ sausage mode is stabilized by
sufficiently diffuse current, while the $|m|\geq 1$ kink is unavoidable
with a free axis.~\cite{Kadomtsev1966}
Such analysis shows the $m=0$ modes to be essentially convective structures having
a marginally stable state, but the $|m|\geq 1$ modes to possess no such marginal state with a free axis.~\cite{Freidberg_2014}

Before reviewing the literature, one observation clarifies the shear-flow-stabilization literature of these MHD modes:
the $m=0$ wavevector is entirely perpendicular to the magnetic field, while $|m|\geq 1$ 
modes have a parallel component.
This distinction is central to the present work,
where the absence or presence of a parallel wavevector component determines
whether the Alfv\'{e}n continuum shields a mode from flow-induced damping,
and thus whether sub-Alfv\'{e}nic or trans-Alfv\'{e}nic shear
is required for stabilization.

With this distinction, the two modes are reviewed in turn.
The $m=0$ instability, interchange, is efficiently suppressed both by diffuse equilibrium
(marginal profiles) and by flow shear.
Indeed, adiabatic flux compression from coaxial accelerators naturally
produces near-marginal profiles~\cite{Marshall1968ContinuousFlowPinch,Crews_2024} 
in which even sub-Alfv\'{e}nic sheared flows suffice to stabilize disruptive interchange modes
arising from non-ideal pinch assembly.~\cite{Meier_2021}
Considering the Bennett pinch, a near-marginal profile, MHD simulations have shown complete interchange 
suppression with modest sheared flows.~\cite{DeSouza_Machado_2000, Paraschiv_2010}
However, strongly super-marginal profiles, such as those driven by vacuum flux compression,
resist interchange stabilization by sheared flow. 
Section~\ref{subsec:m0_stabilization} confirms these results and sheds additional light on them.

Turning to the $m=1$ kink mode, Ref.~\citenum{Shumlak_1995} examined the ideal MHD eigenvalue problem
and reported kink mode suppression with a sub-Alfv\'{e}nic sheared flow.
Reference~\citenum{Arber_1996}, an independent analysis, 
found trans-Alfv\'{e}nic sheared flows were required and, in a subsequent 
Comment,~\cite{Arber_1996_reply} attributed that sub-Alfv\'{e}nic result
to an axis boundary condition artifact;
the authors of Ref.~\citenum{Shumlak_1995} defended their 
numerical treatment in a Reply.~\cite{Shumlak_1996}

The present work confirms the trans-Alfv\'{e}nic condition on shear-stabilization
identified in Ref.~\citenum{Arber_1996} and, using spectral theory, 
clarifies why it arises (see Fig.~\ref{fig:continuum_shielding}).
This same analysis suggests a potential resolution of the apparent contradiction
between the two works, namely that the spectral structure of the problem contains
a proliferation of solution branches that earlier single-mode calculations could
not distinguish.
A near-marginal branch could plausibly have been tracked as the kink mode, leading
to apparent stabilization conditions like $v_z' \geq 0.1 kv_a$ which have not been
reproduced by subsequent independent analyses.

Notably, however, the subjective conclusion of Ref.~\citenum{Arber_1996},
namely that it be ``unlikely [...] a Z‐pinch could in practice be stabilized by the 
introduction of sheared flow,'' was made within the ideal model.
In the context of fluid dynamics, such a conclusion is analogous to 
predicting the impossibility of stable fluid motion from calculations 
at infinite Reynolds number.
Therefore, the present work is merely concerned with a precise understanding 
of the ideal linear dynamics of sheared-flow Z-pinch equilibrium.
Ultimately, questions of stability require higher fidelity models and 
experimental exploration across dimensionless number regimes.

Despite the unsettled theoretical picture,
the ZaP flow Z pinch experiment demonstrated stability for over 700 Alfv\'{e}n times,~\cite{Shumlak_2003, Golingo_2005}
and the subsequent Fusion Z-Pinch Experiment (FuZE) has scaled to higher performance
with sustained thermonuclear neutron production.~\cite{Zhang_2019,Shumlak_2020,Levitt_2024,Ryan_2025}
Yet within ideal MHD, a contemporary ideal eigenmode analysis examining the FuZE experiment 
found that even trans-Alfv\'{e}nic sheared flow fails to stabilize the kink
for any equilibrium profile,~\cite{Angus_2020} in agreement with Ref.~\citenum{Arber_1996}.
It appeared that ideal MHD could not explain the plasma parameters
and lifetime observed by diagnostics.~\cite{Goyon_2024}

Kinetic and multifluid studies were conducted to examine this discrepancy,
in particular in the Hall regime.
For reason of dimensionality, these studies focused on the 
interchange instabilities, which are known to be enhanced in the Hall 
regime for static pinches~\cite{Igitkhanov_Kadomtsev_1970, Sotnikov_2004}
(recall the Hall regime itself occurs at 
low linear density~\cite{Crews_2025}).
These simulations observed robust damping of Hall-regime interchange 
modes with sub-Alfv\'{e}nic shear in both kinetic and multifluid studies,
similar to the trend in ideal MHD but beginning from a 
faster-growing static instability.~\cite{Tummel_2019, Geyko_2021, Meier_2021}
Though these studies illuminated the role of Hall fields 
and kinetic effects in the interchange mode,
what becomes of the kink mode with trans-Alfv\'{e}nic shear, 
and whether the modes that persist should prove
to be MHD instabilities at all, has still not been examined.

The present work resolves this question up to limits of ideal MHD,
noting that the experiments are far from the ideal regime.
Within the ideal model, sheared flow does indeed stabilize 
the MHD kink instability, but the trans-Alfv\'{e}nic flow required 
to do so excites another class of compressible, shear-driven instabilities 
present even without a magnetic field.
It is these modes, not the original MHD instabilities, 
that persist with trans-Alfv\'{e}nic sheared flow.
To go further than this conclusion requires considering non-ideal transport, 
as classical cross-field transport breaks down at the central magnetic null of the Z pinch, 
which is currently under investigation.~\cite{Meier_2021, Crews_2025}

This conclusion has antecedents in astrophysical jet studies
when such are modeled as current-carrying axisymmetric flows. 
Here the distinction between MHD and shear-driven instabilities was drawn long 
ago.~\cite{Cohn1983}
Such studies have since been extended to relativistic 
frameworks.~\cite{Sinnis_2023, Vlahakis_2024}
The present work deepens these studies via spectral analysis.
Note that compressible shear-driven modes emit acoustic energy, and throughout this work ``radiation'' denotes magnetosonic emission into the plasma.

\subsection{Paper organization}
Section~\ref{sec:bvp} formulates the linearized ideal MHD eigenvalue problem
and defines its solutions as the roots of a complex dispersion function $\chi(\omega, k)$.
Section~\ref{sec:spectrum} analyzes the spectrum of the eigenvalue problem,
first reviewing the static MHD instabilities, and then discussing the MHD continuum and its role in shielding or damping,
the adiabatic and resonant parts of the dispersion function in the continuous spectrum, 
marginal modes, and the classification of shear-driven modes.
Section~\ref{sec:results} presents computational results,
covering interchange stabilization and the full dispersion relations of $m=0$ and $m=1$ modes
in trans-Alfv\'{e}nic shear flows.
Section~\ref{sec:conclusion} summarizes the findings and identifies open questions.
Lengthy calculations and background material are contained in three appendices: 
Appendix~\ref{app:regularization} treats regularization of the dispersion function at the continuous spectrum,
Appendix~\ref{app:rayleigh} reviews stability of cylindrical, incompressible shear flows (the classic Kelvin-Helmholtz instability, which is avoided in this work by construction),
and Appendix~\ref{app:wkb} considers compressible acoustics in shear flows.

\section{Ideal linear dynamics of the Z pinch}
\label{sec:bvp}
The governing equations of the ideal MHD model are
\begin{align}
  \frac{d\rho}{dt} &= -\rho\nabla\cdot\vec{v}\label{eq:cont},\\
  \rho\frac{d\vec{v}}{dt} &= -\nabla p + \vec{\jmath}\times\vec{B}\label{eq:moment},\\
  \frac{\partial\vec{B}}{\partial t} &= \nabla\times(\vec{v}\times\vec{B})\label{eq:faraday},\\
  \frac{dp}{dt} &= -\gamma p\nabla\cdot\vec{v}\label{eq:adiabatic_p}
\end{align}
where $\mu_0\vec{\jmath} = \nabla\times\vec{B}$ and $\gamma$ is the specific heat ratio.
Let $\rho_0(r), p_0(r), \vec{B}_0=B_{0\theta}(r)\hat{\theta},$ and $\vec{v}_0=v_{0z}(r)\hat{z}$ be the
density, pressure, magnetic flux density, and radially sheared axial flow of an axially
homogeneous, flowing Z-pinch equilibrium.

Linearizing Eqs.~\ref{eq:cont}-\ref{eq:adiabatic_p} about this equilibrium and Fourier transforming
in the axial $z$, azimuthal $\theta$, and time $t$ coordinates yields normal modes
$\exp(i(kz+m\theta - \omega t))$.
Take the time-transform to be two-sided, \textit{e.g.},
$\hat{f}(\omega)=\int_{-\infty}^\infty e^{-i\omega t}f(t)dt$.
Further, let the perturbed quantities be denoted with a subscript $1$,~\textit{i.e.}, $p_1$, $B_{1\theta}$, etc,
and let the linear displacement vector $\vec{\xi}$ be defined by $\frac{d\vec{\xi}}{dt} \equiv \vec{v}_1$.
Throughout, lengths are normalized to the pinch radius $r_p$, and growth rates and
frequencies to the inverse Alfv\'{e}n time $\tau_A^{-1}$, where $\tau_A \equiv r_p/v_A^*$
is the Alfv\'{e}n transit time built from the reference speed $v_A^*$ defined in
Sec.~\ref{subsec:setup}.

The linearization of Eqs.~\ref{eq:cont}-\ref{eq:adiabatic_p} for the static screw-pinch equilibrium
presented difficulties historically due to the essential and removable singularities of the
so-obtained differential equations.
Hain and L\"{u}st originally cast the problem as a second-order singular differential equation
in $r$,~\cite{Hain_1958} containing singularities at the shear Alfv\'{e}n, slow magnetosonic,
and fast magnetosonic frequencies of a homogeneous equilibrium in a uniform magnetic field.~\cite{Grad_1973}
The fast magnetosonic singularity is apparent and removable, while the shear Alfv\'{e}n
and slow magnetosonic singularities are essential.~\cite{Goedbloed_1998}

Reference~\citenum{Appert_1974} ultimately presented a standard form clearly expressing only the
essential singularities, recasting the dynamics as a first-order problem
\begin{equation}\label{eq:ode_sys}
  \frac{d\vec{u}}{dr}=L\vec{u}
\end{equation}
in terms of the canonical perturbation vector $\vec{u} \equiv \{ r\xi_r,  P\}$ and a self-adjoint, singular
operator $L=L(\omega, k)$, where $\xi_r$ is the radial component of the displacement vector and $P$ the total perturbed pressure
$P \equiv p_1 + \vec{B}_1\cdot\vec{B}_0/\mu_0$.
Reference~\citenum{Bondeson_1987} later observed that this standard form remains
unchanged in the presence of radially sheared axial flow with three modifications:
first, that the frequency $\omega$ be replaced with the Doppler shifted frequency,
$\omega\to \tilde{\omega} \equiv \omega - k v_{0z}(r)$, second, that the displacement vector $\vec{\xi}$
be understood as the displacement in the Lagrangian frame, $\vec{\xi}\to \vec{\xi}_L$, and third, that the
operator $L$ is no longer self-adjoint.

The spectral framework of this article may be extended to azimuthal rotation,~\cite{Frieman_1960, Goedbloed_2018a, Goedbloed_2018b} which enters through Doppler-Coriolis modification of the force operator.
Rotation is omitted in this work because experimentally the sheared-flow Z pinch is driven by axial flow.
However, spectral stability analysis of rotating Z-pinch configurations would be a natural extension of the present analysis.

The linear operator $L$ has the matrix representation
\begin{equation}\label{eq:linear_operator}
  L = \frac{1}{D}\begin{bmatrix} C_1 & -rC_2\\ C_3/r & -C_1\end{bmatrix}
\end{equation}
whose entries are given by
\begin{align}
D &= \left( \rho_0 \tilde{\omega}^2 - \frac{F^2}{\mu_0} \right) \left[ \rho_0 \tilde{\omega}^2 \left( \gamma p_0 + \frac{B_0^2}{\mu_0} \right) - \gamma p_0 \frac{F^2}{\mu_0} \right], \label{eq:denom}\\[1ex]
C_1 &= \frac{2 B_{0\theta}}{\mu_0 r} \left\{ \rho_0^2 \tilde{\omega}^4 B_{0\theta} - \frac{m}{r} F \left[ \rho_0 \tilde{\omega}^2 \left( \gamma p_0 + \frac{B_0^2}{\mu_0} \right) - \gamma p_0 \frac{F^2}{\mu_0} \right] \right\} \\[1ex]
C_2 &= \rho_0^2 \tilde{\omega}^4 - \left( k^2 + \frac{m^2}{r^2} \right) \left[ \rho_0 \tilde{\omega}^2 \left( \gamma p_0 + \frac{B_0^2}{\mu_0} \right) - \gamma p_0 \frac{F^2}{\mu_0} \right] \\[1ex]
C_3 &= D \left[ \rho_0 \tilde{\omega}^2 - \frac{F^2}{\mu_0} + \frac{2 B_{0\theta}}{\mu_0} \frac{d}{dr} \left( \frac{B_{0\theta}}{r} \right) \right] \notag \\
    &\quad + \rho_0 \tilde{\omega}^2 \left( \rho_0 \tilde{\omega}^2 - \frac{F^2}{\mu_0} \right) \left( \frac{2 B_{0\theta}^2}{\mu_0 r} \right)^2 \notag \\
    &\quad - \left[ \gamma p_0 \left( \rho_0 \tilde{\omega}^2 - \frac{F^2}{\mu_0} \right) + \rho_0 \tilde{\omega}^2 \frac{B_{0z}^2}{\mu_0} \right] \left( \frac{2 B_{0\theta} F}{\mu_0 r} \right)^2, \\[1ex]
F &= \vec{k}\cdot\vec{B}_0 = \frac{m}{r} B_{0\theta} + k B_{0z}.
\end{align}
where $B_0=|\vec{B}_0|$. 
This work considers the ideal Z-pinch limit in which $B_{0z}=0$.

Equation~\ref{eq:ode_sys} is cast as a boundary-value problem (BVP) on $r\in [0, r_w]$ by requiring 
$r=0$ to be the axis and $r=r_w$ to be a conducting wall. 
The plasma fills the domain out to this wall.
No vacuum region or asymptotic matching to an external region are considered in this work.
The configurations studied are accordingly internal modes of the equilibrium in the conventional sense.~\cite{Freidberg_2014, Goedbloed_2019}
The explicit axis and wall boundary conditions are given in Eqs.~\ref{eq:axis_bc}--\ref{eq:wall_bc}.
Asymptotic boundary conditions are given in Ref.~\citenum{Appl1992}.
The permissible solutions to this BVP are then expressed by
the roots of an analytic dispersion function $\chi(\omega, k)$, which is obtained as follows.

Denote by $\mathcal{L}(\omega, k)$ the propagator of Eq.~\ref{eq:ode_sys},
\textit{i.e.}, the integral operator that advances solutions $\vec{u}(r)$ from axis to wall,
\begin{equation}\label{eq:integral_operator}
  \vec{u}(r_w) = \mathcal{L}(\omega, k)\vec{u}(0),
\end{equation}
symbolizing the integration of Eq.~\ref{eq:ode_sys} through the plasma.
The axis boundary condition at $r=0$, $\vec{u}(0)$, selects a one-dimensional subspace of 
permissible initial conditions, 
and the conducting wall boundary condition demands that the radial displacement,
a factor of the first component,  vanish at the wall, \textit{i.e.}, $(r\xi_r)|_{r=r_w}=0$.
A functional representation of these boundary conditions applied to the propagator, Eq.~\ref{eq:integral_operator}, is
\begin{equation}\label{eq:dispersion_function}
  \chi(\omega, k) = \begin{bmatrix}1\\ 0\end{bmatrix}^T \cdot \mathcal{L}(\omega, k)\vec{u}(0) = 0
\end{equation}
so that the eigenvalues are the pairs $(\omega, k)$ satisfying Eq.~\ref{eq:dispersion_function}
for each $m$, defining the dispersion function $\chi(\omega, k)$.
Observe that $\chi(\omega, k)$ is simply the value of $r\xi_r$ at the wall for the solution regular at the axis, and its complex zeros are the frequencies where $(r\xi_r)|_{r=r_w}=0$ is satisfied.
A subscript or argument $m$ is not carried for simplicity.

\section{Spectrum of the boundary-value problem}
\label{sec:spectrum}
The spectrum of the Z-pinch equilibrium for given wavevector components $k, m$
has a discrete and a continuous part.
The discrete eigenfrequencies are the complex zeros of the dispersion function
$\chi(\omega,k)$, Eq.~\ref{eq:dispersion_function}, with regular eigenfunctions $\psi_n(r)$,
while the continuous spectrum $\sigma_c$ exists where the governing
ODE, Eq.~\ref{eq:ode_sys}, is singular, meaning it lies on the real-$\omega$ axis.
The continuous spectrum comprises four possibly degenerate sets in
ideal MHD, the Doppler-shifted Alfv\'{e}n and slow magnetosonic resonances, and it supports
singular, improper eigenfunctions $\phi(r;\omega)$.~\cite{Goedbloed_2004, Goedbloed_2019}
An arbitrary solution, therefore, separates into these discrete and continuum parts,
\begin{equation}\label{eq:spectral_decomposition}
  \vec{u}(r,t) = \vec{u}_{\mathrm{d}}(r,t) + \vec{u}_{\mathrm{c}}(r,t),
\end{equation}
with $\vec{u}_{\mathrm{d}}$ assembled from the discrete eigenfunctions $\psi_n$ and
$\vec{u}_{\mathrm{c}}$ from the continuum eigenfunctions $\phi(r;\omega)$.
The singular eigenfunctions of the continuous spectrum encode the transient response to initial data,
analogous to the Case-van Kampen modes in kinetic spectral theory.~\cite{Van_Kampen_1955, Case_1959}
This discrete-plus-continuous structure persists when sheared flow renders the
linearized dynamics non-self-adjoint, whereupon Eq.~\ref{eq:spectral_decomposition}
becomes a bi-orthogonal decomposition, the only difference being its parts need be 
projected out by the adjoint eigenfunctions.~\cite{Frieman_1960, Goedbloed_2019}
The discrete/continuous decomposition is valuable to understand because the 
interaction between discrete modes and the continuum produces continuum damping.
Shear-flow stabilization operates when the flow Doppler-shifts the continuum into resonance
with the instability modes.
The same sheared flow can also drive instabilities through wave resonance.

As the spectral structure captures the how of shear-flow stabilization,
this section discusses the MHD spectral structure of the sheared-flow Z pinch equilibrium.
It begins by reviewing the classic MHD instabilities of the unstable discrete spectrum in Section~\ref{subsec:discrete}.
Section~\ref{subsec:MHD_continuum} then discusses the MHD continuum and continuum damping,
along with the Alfv\'{e}n gap that inhibits shear-flow stabilization of the kink.
Section~\ref{sec:marginal} treats the marginal modes and embedded eigenvalues and their interaction with the continuum,
and finally Section~\ref{subsec:shear_driven} introduces the shear-driven instabilities arising in trans-Alfv\'{e}nic flow.

\subsection{Unstable discrete modes: classic MHD instabilities}\label{subsec:discrete}
This section reviews the disruptive $m=0$ and $m=1$ MHD instabilities of the Z pinch.
The discrete spectrum with $\text{Im}(\omega) \neq 0$ consists of these instabilities for given
components of the wavevector $m$ and $k$, which are further quantized radially.
This section reviews the classic ``stability conditions'' on the equilibrium profiles for MHD 
stability, as derived in Ref.~\citenum{Kadomtsev1966} from the energy principle, which show
a fundamentally different character between the $m=0$ and $m\geq 1$ instabilities.

\subsubsection{Interchange/sausage modes (\texorpdfstring{$m=0$}{m=0})}
The magnetic field of the Z pinch is concave towards high pressure, thereby driving
interchange (or sausage) instability.
However, compressional work by the magnetic field is stabilizing to
the interchange, because the interchange instability is pressure-driven.~\cite{Freidberg2007_Ch12}
In the absence of sheared axial flows, the stabilizing effect outweighs the instability drive
when the equilibrium profile satisfies a condition traditionally given as
\begin{equation}\label{eq:m0_stability}
  -\frac{d\ln p}{d\ln r} \leq \frac{4\gamma}{2 + \gamma\beta}
\end{equation}
where $\beta = p/p_B$ is local plasma beta, $p_B=B_\theta^2/2\mu_0$ is local magnetic pressure,
and $\gamma$ is the ratio of specific heats.~\cite{Kadomtsev1966}
The stability condition, Eq.~\ref{eq:m0_stability}, is equivalent to
\begin{equation}\label{eq:alt_m0_stability}
  \frac{d}{dr}\ln\Big(\frac{p}{\langle j_z\rangle^\gamma}\Big) \leq 0
\end{equation}
where $\langle j_z\rangle = \int j_z dA / \pi r^2$ is area-averaged current density.~\cite{Crews_2024}

To illustrate profile stability to $m=0$ instability, consider the following one-parameter family of diffuse Z-pinch equilibria,
\begin{align}
  p_0/p &= \Big(1 + \frac{\alpha-1}{\alpha}\frac{2}{\beta}\Big)^{\alpha/(\alpha-1)}\label{eq:poly1}\\
  (r/r_p)^2 &= \beta^{-1}\Big(1 + \frac{\alpha-1}{\alpha}\frac{2}{\beta}\Big)^{(2-\alpha)/(\alpha-1)}\label{eq:poly2}
\end{align}
where $p_0$ is on-axis pressure, $r_p$ is characteristic radius, $\beta\in(0,\infty)$ is local beta that 
parameterizes the radial coordinate, and $\alpha\in (1, \infty)$ is a parameter that measures the concentration of electric current.
The case $\alpha=2$ is the Bennett equilibrium
\begin{equation}
  p/p_0 = (1+(r/r_p)^2)^{-2},
\end{equation}
while $\alpha\to\infty$ limits to a uniform current density.

The equilibria have a polytropic property $p/p_0 = (\langle j_z\rangle/j_0)^\alpha$ where $j_0$ is
on-axis current density.
The polytropic property makes the stability condition transparent, as using Eq.~\ref{eq:alt_m0_stability}
shows that interchange stability requires $\alpha \leq \gamma$, 
\textit{i.e.}, sufficiently diffuse current.
The marginal profile $\alpha=\gamma$ is called a Kadomtsev pinch.
Profiles with $\alpha > \gamma$ are interchange unstable, including the Bennett profile for the typical
specific heat ratio $\gamma=5/3$.

\begin{figure}[htbp]
\centering
\includegraphics[width=\columnwidth]{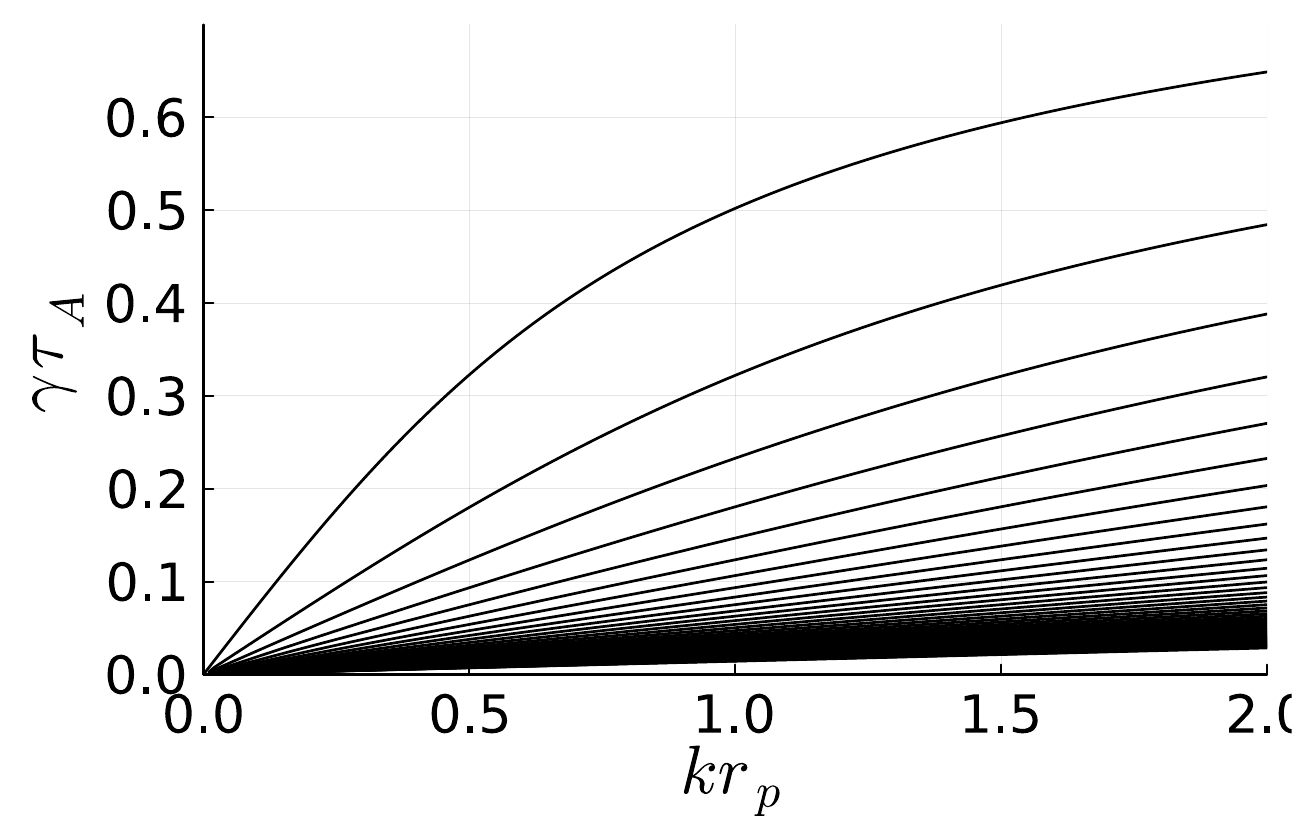}
\caption{Spectrum of the $m=0$ ``sausage'' instability for a static Bennett profile with wall at $r_w = 4r_p$.
The equilibrium index $\alpha=2$ violates $\alpha\leq 5/3$, permitting instability.}
\label{fig:unstable_m0}
\end{figure}

\begin{figure}[htbp]
\centering
\includegraphics[width=\columnwidth]{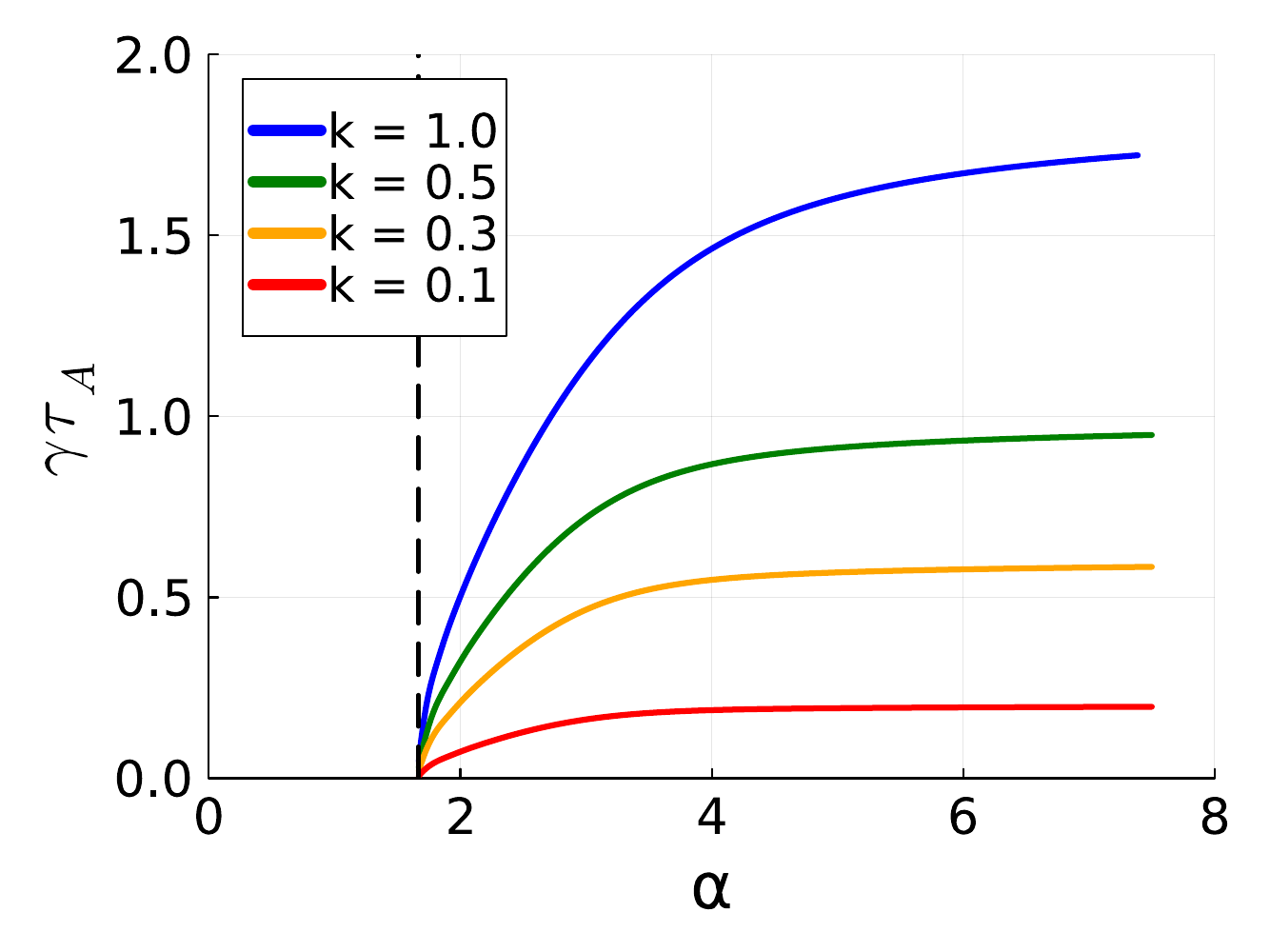}
\caption{Static ($v_0=0$) growth rate of the most unstable $m=0$ mode
as a function of polytropic index $\alpha$ at several axial wavenumbers.
The dashed line marks the Kadomtsev marginal index $\alpha=\gamma=5/3$.
Wall at $r_w = 4r_p$.}
\label{fig:static_growth_vs_alpha}
\end{figure}

The interchange mode is radially quantized, as shown in Fig.~\ref{fig:unstable_m0}
for the Bennett equilibrium ($\alpha=2$), with the first radial mode the fastest growing.
Figure~\ref{fig:static_growth_vs_alpha} shows the growth rate as a function of $\alpha$
at several wavenumbers. The onset of instability at $\alpha = \gamma$ is sudden,
and the growth rate increases monotonically with current concentration.

\subsubsection{Kink modes (\texorpdfstring{$m=1$}{m=1})}
Unlike $m=0$ modes, compressibility offers no stabilization to the kink mode.
This makes the kink the most dangerous mode for confinement, and it is often the one that causes disruptions.
This is not to say the kink mode is insensitive to the equilibrium profile.
In fact, from the energy principle, the stability condition for $m\geq 1$ in the Z pinch is~\cite{Kadomtsev1966}
\begin{equation}\label{eq:m1_stability}
  -\frac{d\ln p}{d\ln r} \leq \frac{m^2}{\beta},
\end{equation}
which for $m=1$, combined with force balance, takes the form
\begin{equation}\label{eq:alt_m1_stability}
  \frac{d}{dr}(rp_B) < 0.
\end{equation}
Since $rp_B\to 0$ as $r\to 0$ and $rp_B>0$ for $r>0$, the derivative must be positive somewhere,
so no profile with a free axis is kink stable.~\cite{Freidberg2007_Ch12}
Notably, a so-called ``hard-core'' profile $p_B\sim 1/r$ surrounding a rigid conductor is kink marginal,
but the local $\beta$ cannot exceed unity in such a case.

The profile effects can be understood by casting 
the stability condition into one based on plasma $\beta$.
This follows from integrating Eq.~\ref{eq:alt_m1_stability} in the alternative form
\begin{equation}\label{eq:diff_inq}
  -\frac{dp}{dr} \leq -\frac{dp_B}{dr}.
\end{equation}
When $\beta(r)$ decreases monotonically in the absence of background pressure,
Eq.~\ref{eq:diff_inq} integrates to
\begin{equation}\label{eq:critical_beta}
  \beta \leq \beta^*,
\end{equation}
where the critical beta $\beta^* \equiv \beta(r^*)$ is the local beta at the marginal surface
$r=r^*$ where $dp/dr = dp_B/dr$.
It is thus the high-$\beta$ core that is actually unstable to the kink.
For the Bennett equilibrium, $\beta^*=1/3$ with the unstable core
extending to $r^* = \sqrt{3}r_p$.
In a sharp pinch (\textit{i.e.}, axial surface current separating pressurized plasma from an external vacuum), 
$\beta^*= 0$ so that the entire plasma column contributes to instability.
The marginal ``hard-core'' profiles have $\beta^* = 1$.

\begin{figure}[htbp]
\centering
\includegraphics[width=\columnwidth]{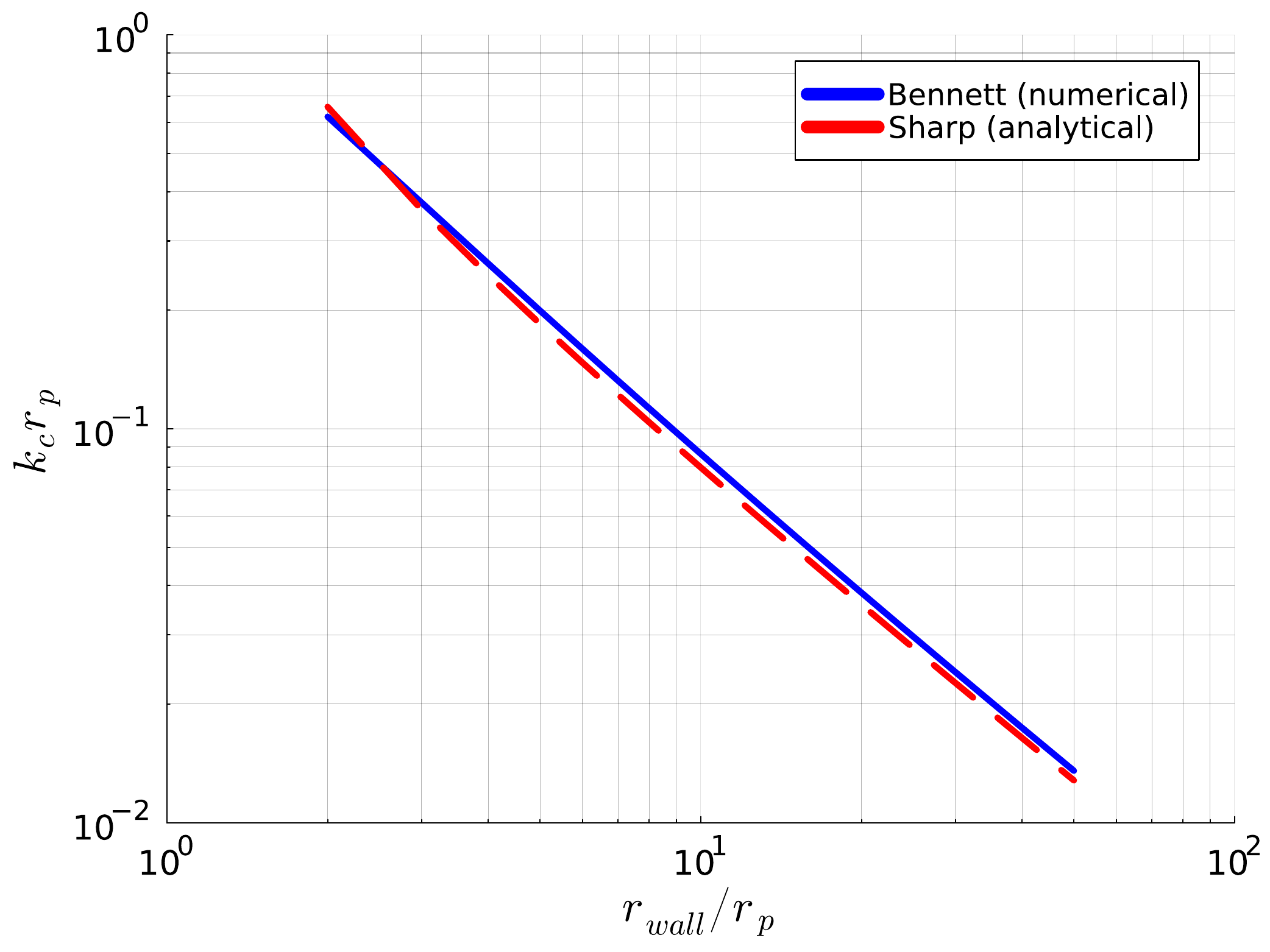}
\caption{Critical wavenumber $k_c$ for wall stabilization of the $m=1$ kink mode
as a function of wall radius.
The sharp pinch result (Eq.~\ref{eq:kink_wall_stabilization}) closely matches the diffuse Bennett profile.}
\label{fig:wall_stabilization}
\end{figure}

\begin{figure}[htbp]
\centering
\includegraphics[width=\columnwidth]{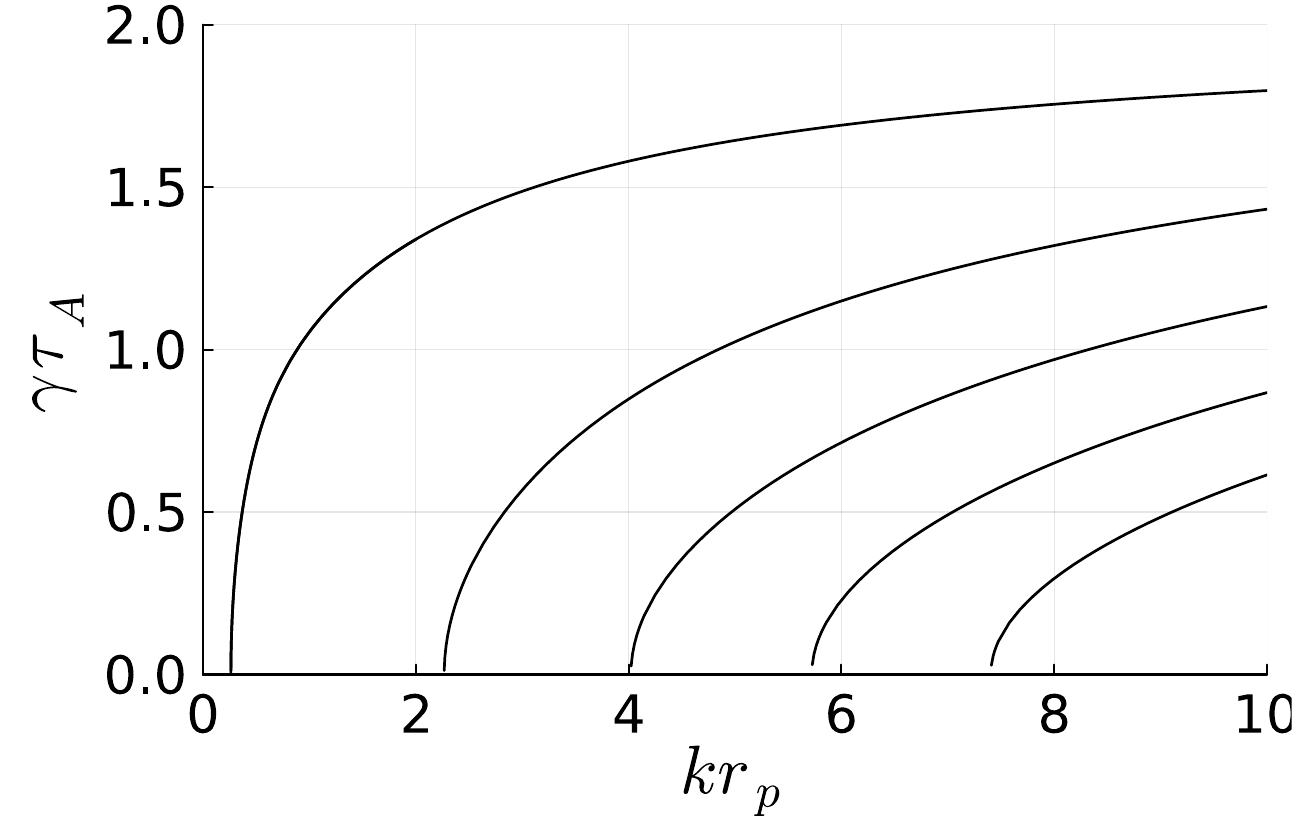}
\caption{Instability spectrum of the $m=1$ kink instability for a static Bennett profile with wall at $r_w = 4r_p$.}
\label{fig:unstable_m1}
\end{figure}

A perfectly conducting wall located at $r=r_w$ stabilizes long-wavelength kinks under a critical wavenumber, $k<k_c$, 
similar to external kink stabilization of screw-pinch configurations.~\cite{Freidberg_2014}
The incompressible linear response of a sharp pinch 
excellently approximates the critical wavenumber for wall stabilization of diffuse profiles.
In this approximation, $k_c$ satisfies the transcendental equation
\begin{equation}\label{eq:kink_wall_stabilization}
  1 + kr_p\Big(\frac{I_w' K_p - I_p K_w'}{I_w' K_p' - I_p' K_w'}\Big) = 0,
\end{equation} 
where $I_p \equiv I_1(kr_p)$, $K_w' \equiv K_1'(kr_w)$, etc., with $I_1$ and $K_1$ modified Bessel functions
of the first and second kind.~\cite{Kadomtsev1966, Freidberg_2014}
Figure~\ref{fig:wall_stabilization} compares the critical wavenumber from Eq.~\ref{eq:kink_wall_stabilization}
with numerical results for Bennett equilibrium truncated at $r=r_w$ by a conducting wall; 
both follow $k_c r_p \sim (r_w/r_p)^{-1.1}$ asymptotically.
Figure~\ref{fig:unstable_m1} plots the Bennett equilibrium's unstable $m=1$ spectrum for $r_w=4r_p$,
showing that the kink mode is unstable for essentially all wavenumbers.

\subsection{The MHD continuum: resonance and phase mixing}\label{subsec:MHD_continuum}
Having reviewed the discrete modes, this review turns to the MHD continuum.
The continuous spectrum arises from local resonances between the linear perturbation and the natural frequencies of the equilibrium.
The set of all such natural frequencies across the plasma column comprises the continuous spectrum.
Global modes interact with local modes through these resonances, leading to damping or excitation.
Section~\ref{subsubsec:continuous_spectrum} examines the effect of sheared axial flow on the basic MHD continuum,
and Section~\ref{subsubsec:phase_mixing} discusses phase mixing effects on the $m=0$ and $m=1$ instability modes.

\subsubsection{Doppler-shifted resonances and continuous spectrum}\label{subsubsec:continuous_spectrum}
The zeros of the denominator $D$, Eq.~\ref{eq:denom}, determine the linear resonances of the Z-pinch equilibrium.
Factoring out $\rho_0^2(\gamma p_0 + B_0^2/\mu_0)\neq 0$, the root-containing factors are
\begin{equation}\label{eq:denominator_form}
  D \sim (\tilde{\omega}^2 - m^2\omega_a^2)(\tilde{\omega}^2 - m^2\omega_{ms}^2).
\end{equation}
Therefore, resonance occurs at Doppler-shifted $m$-multiples of the
Alfv\'{e}n and slow magnetosonic frequencies, namely
\begin{align}
  \omega_a^2 &\equiv \frac{v_a^2}{r^2},\\
  \omega_{ms}^2 &\equiv \frac{1}{r^2}\frac{c_s^2v_a^2}{c_s^2 + v_a^2}.
\end{align}
Here, $v_a = |\vec{B}_0|/\sqrt{\mu_0\rho_0}$ and $c_s = \sqrt{\gamma p_0/\rho_0}$ 
are the local Alfv\'{e}n and sonic speeds.
Namely, linear resonance occurs at frequencies $\omega^*$ in the pairs
\begin{equation}\label{eq:resonant_frequencies}
  \omega^* = \{kv_{0z} \pm m\omega_a, kv_{0z} \pm m\omega_{ms}\}
\end{equation}
which depend on radius through the flow profile $v_{0z}=v_{0z}(r)$ and the 
equilibrium density, temperature, and magnetic field profiles.
The continuous spectrum consists of $\omega^*$ over all radii,
\begin{equation}
  \sigma_c = \bigcup_{r\in[0,r_w]} \{\omega^*(r)\} = \sum_s \bigcup_{r\in[0,r_w]} \{\omega_s^*(r)\}
\end{equation}
with $s$ describing the four branches in Eq.~\ref{eq:resonant_frequencies}.
The continuum eigenfunctions $\phi_s(r; \omega)$ are singular at the resonant surface $r=r_c$
where $\omega = \omega^*_s(r_c)$.
These improper eigenfunctions are non-square integrable and exhibit both logarithmic and
step discontinuities at the resonant layer.~\cite{Case_1960, Goedbloed_2004}

The resonance conditions of Eq.~\ref{eq:resonant_frequencies} have several notable and 
immediate consequences.
First, each radial shell of the plasma possesses four generally radially sheared and 
possibly degenerate natural frequencies $\omega = \omega^*(r)$ depending on the 
equilibrium profiles of density, temperature, and magnetic flux and from the Doppler effect of 
radially sheared axial flow.
When $d\omega^*/dr \neq 0$, adjacent radial shells resonate at slightly different 
frequencies, leading to phase mixing. 

Secondly, an eigenmode,~\textit{i.e.}, a \textit{global} mode, at frequency $\omega$ either resonantly 
deposits or extracts energy at the \textit{local} radius $r_c$ where $\omega=\omega^*(r_c)$.
The interplay of local-global resonant interaction with phase mixing of the 
local modes is well-known in plasma physics from the linear kinetic theory
of plasma oscillations and the paradigm of Landau damping, and
shear-flow-induced continuum damping is analogous.

\subsubsection{Phase mixing and spectral accessibility}\label{subsubsec:phase_mixing}
This section examines phase mixing by the MHD continuum and its dependence on the perturbation's spectral structure.
The key concept is spectral accessibility, meaning that for continuum damping to occur,
the perturbation's frequency spectrum must overlap with resonant surfaces in the plasma.

Considering a perturbation of radial extent $\delta r$ centered on a resonant surface,
each radius oscillates at its local natural frequency $\omega^*(r)$,
accumulating a phase difference $\Delta\phi\sim |d\omega^*/dr|\delta r \Delta t$ in its 
continuous spectrum over time $\Delta t$.
As these phase differences accumulate, the initial perturbation is forgotten
and the continuum contribution to the perturbation vanishes,
formally $\lim_{t\to\infty}\int_{\sigma_c} A(\omega)\phi(r;\omega)e^{-i\omega t}d\omega \to 0$
(the Riemann-Lebesgue lemma)
where $A(\omega)$ are amplitudes of the perturbation's projection onto the continuous spectrum.~\cite{Tataronis_1973}
Physically, the perturbation's energy cascades to small scales.
The key point is that while the phase mixing rate is controlled by $|d\omega^*/dr|$, 
to be mixed at all the perturbation's spectrum must lie in a region of $(\omega, k)$-space covered by resonances.

\begin{figure}[htbp]
\centering
\begin{tikzpicture}
\begin{axis}[
    name=m0plot,
    width=\columnwidth,
    height=0.55\columnwidth,
    xlabel={$kr_p$},
    ylabel={$\omega/\omega_a$},
    title={(a) $m=0$: zero-frequency continuum, $\tilde{\omega}=0$},
    xmin=0, xmax=4,
    ymin=-0.25, ymax=2.2,
    axis lines=middle,
    clip=true,
    ytick={1, 2},
    yticklabels={$0.5$, $1$},
]
\addplot[blue!100!red, thin, domain=0:3.6, samples=2] {tanh(0.1^2) * x};
\addplot[blue!95!red, thin, domain=0:3.6, samples=2] {tanh(0.2^2) * x};
\addplot[blue!90!red, thin, domain=0:3.6, samples=2] {tanh(0.3^2) * x};
\addplot[blue!85!red, thin, domain=0:3.6, samples=2] {tanh(0.4^2) * x};
\addplot[blue!80!red, thin, domain=0:3.6, samples=2] {tanh(0.5^2) * x};
\addplot[blue!70!red, thin, domain=0:3.6, samples=2] {tanh(0.6^2) * x};
\addplot[blue!60!red, thin, domain=0:3.6, samples=2] {tanh(0.7^2) * x};
\addplot[blue!50!red, thin, domain=0:3.6, samples=2] {tanh(0.8^2) * x};
\addplot[blue!40!red, thin, domain=0:3.6, samples=2] {tanh(0.9^2) * x};
\addplot[blue!30!red, thin, domain=0:3.6, samples=2] {tanh(1.0^2) * x};
\addplot[blue!25!red, thin, domain=0:3.6, samples=2] {tanh(1.1^2) * x};
\addplot[blue!20!red, thin, domain=0:3.6, samples=2] {tanh(1.2^2) * x};
\addplot[blue!15!red, thin, domain=0:3.6, samples=2] {tanh(1.3^2) * x};
\addplot[blue!10!red, thin, domain=0:3.6, samples=2] {tanh(1.5^2) * x};
\addplot[blue!5!red, thin, domain=0:3.6, samples=2] {tanh(1.7^2) * x};
\addplot[blue!0!red, thin, domain=0:3.6, samples=2] {tanh(2.0^2) * x};
\node[blue, anchor=west] at (axis cs:0.2,-0.15) {\footnotesize $r\to 0$};
\node[red, anchor=west] at (axis cs:0.1,0.75) {\footnotesize $r\to r_w$};
\end{axis}

\begin{axis}[
    at={(m0plot.south)},
    anchor=north,
    yshift=-1.5cm,
    width=\columnwidth,
    height=0.55\columnwidth,
    xlabel={$kr_p$},
    ylabel={$\omega/\omega_a$},
    title={(b) $m=1$: Alfv\'{e}nic continuum $\tilde{\omega}=\pm\omega_a$},
    xmin=0, xmax=4,
    ymin=-2.4, ymax=4.0,
    axis lines=middle,
    clip=true,
    ytick={-2, 0, 2},
    yticklabels={$-1$, $0$, $1$},
]
\addplot[blue!100!red, thin, domain=0:3.6, samples=2] {tanh(0.1^2) * x + 2.0};
\addplot[blue!95!red, thin, domain=0:3.6, samples=2] {tanh(0.2^2) * x + 2.0};
\addplot[blue!90!red, thin, domain=0:3.6, samples=2] {tanh(0.3^2) * x + 2.0};
\addplot[blue!85!red, thin, domain=0:3.6, samples=2] {tanh(0.4^2) * x + 2.0};
\addplot[blue!80!red, thin, domain=0:3.6, samples=2] {tanh(0.5^2) * x + 2.0};
\addplot[blue!70!red, thin, domain=0:3.6, samples=2] {tanh(0.6^2) * x + 2.0};
\addplot[blue!60!red, thin, domain=0:3.6, samples=2] {tanh(0.7^2) * x + 2.0};
\addplot[blue!50!red, thin, domain=0:3.6, samples=2] {tanh(0.8^2) * x + 2.0};
\addplot[blue!40!red, thin, domain=0:3.6, samples=2] {tanh(0.9^2) * x + 2.0};
\addplot[blue!30!red, thin, domain=0:3.6, samples=2] {tanh(1.0^2) * x + 2.0};
\addplot[blue!25!red, thin, domain=0:3.6, samples=2] {tanh(1.1^2) * x + 2.0};
\addplot[blue!20!red, thin, domain=0:3.6, samples=2] {tanh(1.2^2) * x + 2.0};
\addplot[blue!15!red, thin, domain=0:3.6, samples=2] {tanh(1.3^2) * x + 2.0};
\addplot[blue!10!red, thin, domain=0:3.6, samples=2] {tanh(1.5^2) * x + 2.0};
\addplot[blue!5!red, thin, domain=0:3.6, samples=2] {tanh(1.7^2) * x + 2.0};
\addplot[blue!0!red, thin, domain=0:3.6, samples=2] {tanh(2.0^2) * x + 2.0};
\addplot[blue!100!red, thin, domain=0:3.6, samples=2] {tanh(0.1^2) * x - 2.0};
\addplot[blue!95!red, thin, domain=0:3.6, samples=2] {tanh(0.2^2) * x - 2.0};
\addplot[blue!90!red, thin, domain=0:3.6, samples=2] {tanh(0.3^2) * x - 2.0};
\addplot[blue!85!red, thin, domain=0:3.6, samples=2] {tanh(0.4^2) * x - 2.0};
\addplot[blue!80!red, thin, domain=0:3.6, samples=2] {tanh(0.5^2) * x - 2.0};
\addplot[blue!70!red, thin, domain=0:3.6, samples=2] {tanh(0.6^2) * x - 2.0};
\addplot[blue!60!red, thin, domain=0:3.6, samples=2] {tanh(0.7^2) * x - 2.0};
\addplot[blue!50!red, thin, domain=0:3.6, samples=2] {tanh(0.8^2) * x - 2.0};
\addplot[blue!40!red, thin, domain=0:3.6, samples=2] {tanh(0.9^2) * x - 2.0};
\addplot[blue!30!red, thin, domain=0:3.6, samples=2] {tanh(1.0^2) * x - 2.0};
\addplot[blue!25!red, thin, domain=0:3.6, samples=2] {tanh(1.1^2) * x - 2.0};
\addplot[blue!20!red, thin, domain=0:3.6, samples=2] {tanh(1.2^2) * x - 2.0};
\addplot[blue!15!red, thin, domain=0:3.6, samples=2] {tanh(1.3^2) * x - 2.0};
\addplot[blue!10!red, thin, domain=0:3.6, samples=2] {tanh(1.5^2) * x - 2.0};
\addplot[blue!5!red, thin, domain=0:3.6, samples=2] {tanh(1.7^2) * x - 2.0};
\addplot[blue!0!red, thin, domain=0:3.6, samples=2] {tanh(2.0^2) * x - 2.0};
\fill[gray, opacity=0.35] (axis cs:0,-2.0) rectangle (axis cs:3.6,2.0);
\node at (axis cs:1.0,0.975) {\small Alfv\'{e}nic frequency band};
\node at (axis cs:0.8,0.5) {\small $|\omega|<\omega_a$};
\end{axis}
\end{tikzpicture}
\caption{Continuous spectrum in $(\omega, k)$ space shown as Doppler shifted resonances for 
(a) $m=0$ and (b) $m=1$, with frequencies normalized to the Alfv\'{e}n frequency $\omega_a$
so the band edges lie at $\pm1$, and $v_{0z}(r) = \tanh(r^2)$. %
Each line indicates a resonance $\omega = kv_{0z} \pm m\omega_a$ at a
critical radius colored from axis (blue) to wall (red).
The $k=0$ offsets at $\omega=\pm\omega_a$ for $m=1$ create an unmixed gap in the 
Alfv\'{e}nic frequency band (shaded) around $k=0$, shielding the low-frequency, long-wavelength 
spectrum from phase mixing.}
\label{fig:continuum_shielding}
\end{figure}
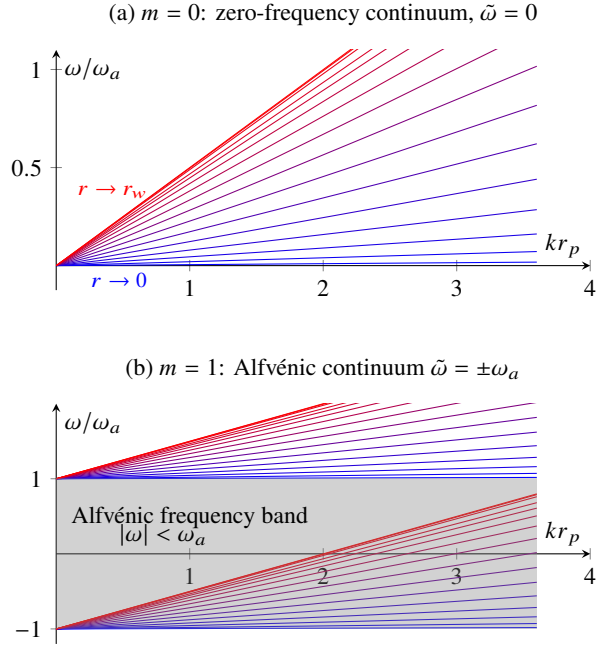

The notion of spectral accessibility is that continuum damping occurs only when the instability couples 
to the continuum at resonant surfaces,
and this constitutes an essential difference in the continuum damping of $m=0$ and $m=1$ perturbations.
The only resonance for axisymmetric perturbations ($m=0$) is $\omega^*=kv_{0z}(r)$.
The continuum lines in $(\omega, k)$-space fan out from the origin, as shown in Fig.~\ref{fig:continuum_shielding}a, 
thereby phase mixing of low-frequency, long-wavelength perturbations at all radii.
This resonance is described by a second-order pole in $\chi(\omega, k)$ for $m=0$.

On the other hand, the resonances for non-axisymmetric, $|m|>0$ perturbations occur at
$\omega^* = kv_{0z}(r) \pm m\omega_a$ and $\omega^* = kv_{0z}(r) \pm m\omega_{ms}$,
described by four simple poles in $\chi(\omega, k)$ for $m\neq 0$.
The $m=1$ shear Alfv\'{e}n resonances provide a clear example,
for which the continuum lines Doppler-shifted by sheared flow originate at the 
Alfv\'{e}n frequencies $\pm\omega_a$, forming a gap in the band $|\omega|<\omega_a$, 
as illustrated in Fig.~\ref{fig:continuum_shielding}b.

\subsubsection{An ``Alfv\'{e}n gap'' shields kink modes from the continuum}
Within the gap $|\omega|< m\omega_a$ there is no continuum damping.
Thus, flow-shear stabilization differs qualitatively for $m=0$ and $|m|>0$ instabilities.
While $m=0$ modes are affected at long-wavelength by
any magnitude of flow shear, this ``Alfv\'{e}n gap'' inhibits continuum damping of $m=1$ instabilities.
For $m=1$ modes to be continuum damped, sheared flow must Doppler-shift the 
counter-flow-propagating shear Alfv\'{e}n waves to propagate co-flow, 
that they may resonate with the kink.
Additionally, to interact in a significant way with the kink mode, the resonant surface must 
lie within the Z-pinch core.
A simple criterion on the magnitude of flow shear necessary for kink modes to be continuum-damped by flow shear then follows from such requirements.

The criterion is estimated as follows for an isothermal Bennett equilibrium, 
for which the Alfv\'{e}n frequency is radially constant, $\omega_a = v_a(r_p)/r_p$,
where $v_a(r_p)$ is the Alfv\'{e}n speed at the pinch radius.
The lower resonance $\omega = kv_{0z} - \omega_a$ interacts strongly with the kink mode when it coincides with the kink frequency $\omega_{kink}$ and the resonant surface lies at $r\leq r_p$.
Further, the kink frequency is bounded by the Alfv\'{e}n frequency as the flow Doppler-shifts it from $\omega=0$ into the sub-Alfv\'{e}nic band, $\omega_{kink} \in [0, \omega_a]$.

Considering then a sheared flow $v_{0z} = v_0 f(r)$ where $f(r)$ is a monotonic shape function with
$f(0)=0$ and $f(r_p)=1$, so that $v_0 \equiv v_{0z}(r_p)$ is the flow speed at the pinch radius,
the resonance reaches the kink frequency at $r=r_p$ when
\begin{equation}
  v_0 \, kr_p \geq (\omega_{kink}/\omega_a + 1)\, v_a(r_p).
\end{equation}
For long-wavelength modes $kr_p \lesssim 1$,
\begin{equation}\label{eq:damping_criterion}
  v_0 \gtrsim v_a(r_p),
\end{equation}
so sheared axial flow must at least transition through the Alfv\'{e}nic speed across the
pinch radius to damp long-wavelength kink modes in the ideal Z pinch.
This criterion is independent of the specific shape function $f(r)$, 
depending only on its endpoints.
Equation~\ref{eq:damping_criterion} is a criterion for the on-set of shear-flow stabilization,
and the \textit{particular character} of the resonant interaction depends on the flow's shape function.

The type of continuum damping discussed in Section~\ref{subsubsec:phase_mixing} 
in fact describes phase mixing of initial perturbations.
Interaction between discrete modes and the continuum is mediated by the
embedded eigenvalues, which is addressed next.

\subsection{Marginal modes and continuum interaction}\label{sec:marginal}
For real $\omega$, the integration through the plasma %
may encounter a resonant surface where the denominator $D$ vanishes, \textit{i.e.}, a pole on the real-$r$ axis.
This integration is, of course, undefined.
The classical approach to treat such resonances, developed by Newcomb for the energy principle,
integrates from both the axis and the wall to the resonant surface and matches the solutions using
Frobenius-derived connection formulae.~\cite{Newcomb_1960, Glasser_2016, Glasser_2025}

This work considers the dispersion function rather than the energy functional, 
so a different approach is taken, one better suited to computing eigenvalues.
The integration path is taken directly through the resonant layer and the singular integral 
is interpreted via its principal value.
This prescription permits both calculation of the embedded eigenvalues and measurement of the
continuum's interaction with discrete global modes.

This section presents the main results, while the mathematical details are given in Appendix~\ref{app:regularization}.
First, Section~\ref{subsubsec:regularization} describes regularization of the problem for $\text{Im}(\omega)=0$,
defining the adiabatic and resonant dispersion functions.
Section~\ref{subsubsec:embedded} introduces the embedded eigenvalues as roots of the adiabatic dispersion function.
Then, Section~\ref{subsubsec:continuum_damping} applies the adiabatic and resonant dispersion functions 
to describe the resonant damping and excitation of global modes by continuum coupling.

\subsubsection{Adiabatic and resonant dispersion functions}\label{subsubsec:regularization}
Consider the limits $\text{Im}(\omega)\to 0^\pm$ from above and below the real-$\omega$ axis
and define functions $\chi^+(\omega_r, k)\equiv \chi(\omega_r + i0, k)$ and
$\chi^-(\omega_r, k)\equiv \chi(\omega_r - i0, k)$.
These functions differ by residue contributions from poles approaching the real-$r$ axis.~\cite{Krall1973, Nicholson1983}
The original ideal MHD system, Eqs.~\ref{eq:cont}-\ref{eq:adiabatic_p}, is real-valued,
so these two limits $\chi^\pm$ are complex conjugates, $\chi^- = (\chi^+)^*$.

The difference of these two functions,
\begin{equation}\label{eq:resonant_dispersion}
  \chi^+ - \chi^- = 2i\text{Im}[\chi^+],
\end{equation}
which is purely imaginary, encodes resonant interaction between global modes and the continuum, 
as discussed in the following Section~\ref{subsubsec:continuum_damping}.
Furthermore, the purely real sum
\begin{equation}\label{eq:regularized_dispersion}
  \chi^+ + \chi^- = 2\text{Re}[\chi^+]
\end{equation}
gives the principal part of integration through a resonant surface,
thereby measuring the adiabatic interaction.
Equations~\ref{eq:resonant_dispersion} and~\ref{eq:regularized_dispersion} identify a physically meaningful treatment
of $\text{Im}(\omega)=0$ in the continuous spectrum, namely an adiabatic dispersion function $\chi_{\text{adi}} \equiv \text{Re}[\chi^+]$
and a resonant function $\chi_{\text{res}} \equiv \text{Im}[\chi^+]$.

\subsubsection{Embedded eigenvalues: mediating continuum interaction}\label{subsubsec:embedded}
\begin{figure}[htbp]
\centering
\begin{tikzpicture}
  \def\xmin{1.716855}
  \def\xmax{2.282540}
  \def\scale{13.435029}  %
  \def\omegaa{2.0}       %

  \fill[gray!30]
    ({(\xmin-\xmin)*\scale}, -0.35) rectangle ({(\omegaa-\xmin)*\scale}, 0.35);

  \draw[thick, <->] (-0.3, 0) -- ({(\xmax-\xmin)*\scale + 0.3}, 0);

  \draw[red!70!black, thick] ({(\omegaa-\xmin)*\scale}, -0.45) -- ({(\omegaa-\xmin)*\scale}, 0.45);

  \foreach \freq in {1.8147, 1.9208, 1.9550, 1.9708, 1.9882, 2.0003, 2.0106, 2.0222, 2.0352, 2.0526, 2.0674, 2.0908, 2.1309, 2.2127} {
    \draw[blue, thick] ({(\freq-\xmin)*\scale-0.12}, -0.18) -- ({(\freq-\xmin)*\scale+0.12}, 0.18);
    \draw[blue, thick] ({(\freq-\xmin)*\scale-0.12}, 0.18) -- ({(\freq-\xmin)*\scale+0.12}, -0.18);
  }

  \foreach \tick in {1.8, 1.9, 2.0, 2.1, 2.2} {
    \draw ({(\tick-\xmin)*\scale}, -0.15) -- ({(\tick-\xmin)*\scale}, 0.15);
    \node[below] at ({(\tick-\xmin)*\scale}, -0.2) {\footnotesize \tick};
  }

  \node[above] at ({(\xmax-\xmin)*\scale}, 0.1) {$\omega\,\tau_A$};
\end{tikzpicture}
\caption{Accumulation of $m=1$, $kr_p=1$ marginal modes at the Alfv\'{e}n frequency of an isothermal, static Bennett profile.
The gray region is the slow magnetosonic continuous spectrum,
and the red line the Alfv\'{e}n frequency $\omega_a = 2\,v_A^*/r_p$.
Modes accumulate to $\omega_a$ from both sides: solutions in the continuum are embedded eigenvalues,
and those outside are non-singular.}
\label{fig:spectral_structure}

\vspace{\floatsep}
\centering
\includegraphics[width=\columnwidth]{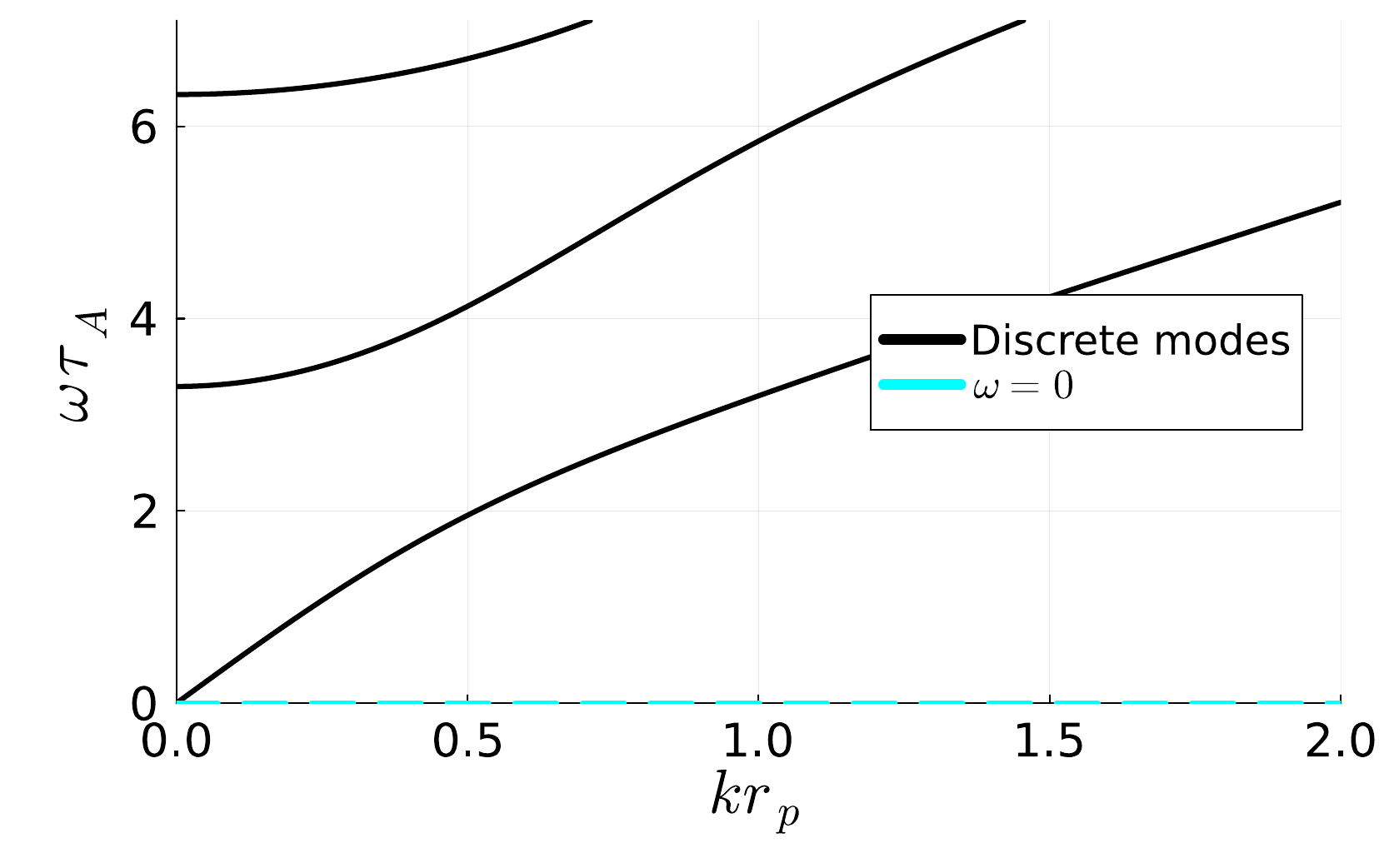}
\caption{Dispersion relation of $m=0$ marginal modes for a static Bennett equilibrium with wall at $r_w = 4r_p$,
showing perpendicular-propagating fast magnetosonic waves.}
\label{fig:marginal_dispersion_m0}

\vspace{\floatsep}
\centering
\includegraphics[width=\columnwidth]{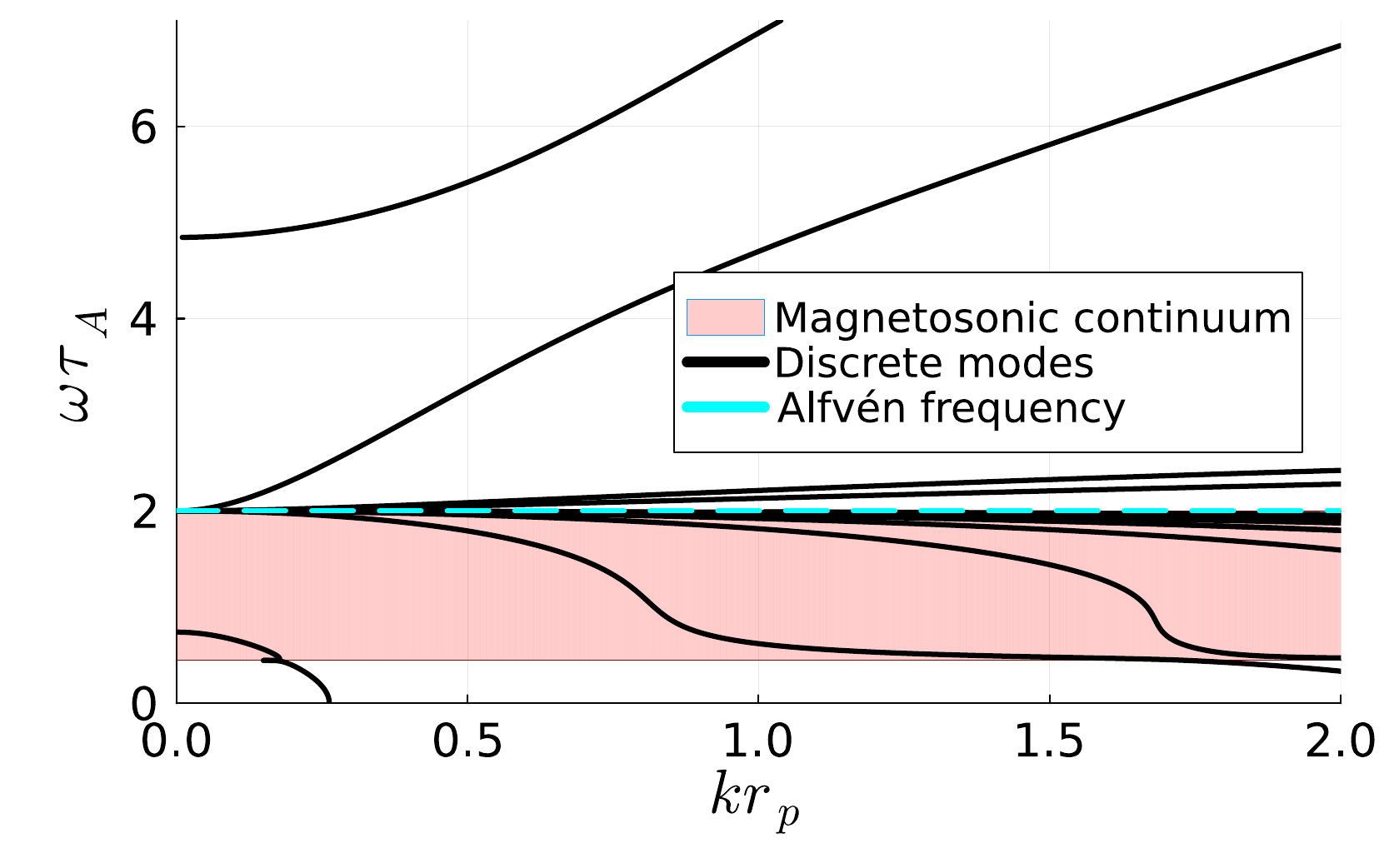}
\caption{Marginal mode dispersion relation for $m=1$ perturbations of a static Bennett profile
with wall at $r_w = 4r_p$.
The shaded region in red is the slow magnetosonic continuum, bounded above by the Alfv\'{e}n frequency (dashed).
Discrete modes thread through the continuum as embedded eigenvalues and emerge as non-singular modes outside.}
\label{fig:marginal_dispersion_m1}
\end{figure}

The roots of $\chi_{\text{adi}}$ are the embedded eigenvalues, components of the discrete spectrum that lie within
the continuous spectrum and whose eigenfunctions are singular on their resonant surfaces.
Unlike the continuum modes, which are in fact ``improper'' solutions of an associated inhomogeneous BVP 
(thereby propagating the initial-value problem, and thus constituting parts of the Green's function),~\cite{Case_1960}
the embedded eigenvalues solve the homogeneous BVP through the adiabatic part of the resonant interaction.
These are marginal modes in local resonance with the continuum
and in global conformity with the boundary conditions,
thereby exemplifying the local-global coupling.
Both embedded and non-embedded marginal modes accumulate at the continuum edge, shown in
Fig.~\ref{fig:spectral_structure}.
The two-sided accumulation toward the Alfv\'{e}n frequency illustrates the oscillation theorem of static ideal MHD spectra,~\cite{Goedbloed_1974} the accumulation of the non-embedded modes in particular being treated in Ref.~\citenum{Goedbloed_2004}.

In static MHD equilibrium, it is well known that $\omega^2$ is real.
The $m=0$ modes have no resonance other than at zero,
so the marginal mode spectrum merely consists of radially quantized fast magnetosonic waves
(Fig.~\ref{fig:marginal_dispersion_m0}).
The $m=1$ modes, by contrast, have resonances at the Alfv\'{e}n and slow magnetosonic frequencies,
and the marginal mode spectrum reflects this richer structure, containing discrete modes 
within the slow magnetosonic continuum as embedded eigenvalues and above the Alfv\'{e}n 
frequency as non-singular modes, shown in Fig.~\ref{fig:marginal_dispersion_m1}.

\subsubsection{Continuum resonance: damping and excitation}\label{subsubsec:continuum_damping}
The embedded eigenvalues are marginally stable, but the continuum they inhabit mediates
energy transfer with discrete modes.
When an unstable mode's frequency is near the continuum, resonant interaction can damp it
by transferring energy to the continuum.
Conversely, in the presence of sheared flow, embedded eigenvalues of opposite energy sign
can couple through the continuum and drive instabilities.~\cite{Hirota_2008}

The effect on discrete modes of resonant interaction with the continuum is measured by
expanding the dispersion function for small imaginary parts.~\cite{Krall1973}
To handle the branch cut in $\chi(\omega)$ on the real-$\omega$ axis, 
consider a small $\text{Im}(\omega)\equiv \gamma>0$ in the upper-half-$\omega$ plane.
The choice of half-plane is only convenient, and the same result is found by considering 
the lower-half-$\omega$ plane.
Expanding gives
\begin{equation}\label{eq:chi_expansion}
  \chi(\omega_r + i0 + i\gamma) = \chi^+\big|_{\omega_r} + i\gamma\partial_{\omega_r}\chi^+\big|_{\omega_r} + \mathcal{O}(\gamma^2).
\end{equation}
The Cauchy-Riemann equations provide the second term in Eq.~\ref{eq:chi_expansion} since $\chi^+$ is analytic 
in its half-plane.
Separating the real and imaginary parts by $\chi^+ = \chi_\text{adi} + i\chi_\text{res}$ gives
\begin{equation}
  \chi(\omega_r + i0 + i\gamma) = \chi_{\text{adi}}- \gamma\chi_{\text{res}}^\prime + i(\chi_{\text{res}} + \gamma\chi_{\text{adi}}^\prime) + \mathcal{O}(\gamma^2)
\end{equation}
where all functions are evaluated at $\omega_r$.
Putting $\text{Im}[\chi] = 0$ (assuming $\omega=\omega_r+i\gamma$ is close to $\chi(\omega) = 0$) suggests 
\begin{equation}\label{eq:continuum_damping}
   \gamma = -\chi_{\text{res}}/\chi_{\text{adi}}^\prime
\end{equation}
for the continuum damping rate near an eigenvalue.
This is a standard expression in the theory of continuum damping.~\cite{Krall1973, Tataronis_1975}

Equation~\ref{eq:continuum_damping} is an incomplete description, however, of the continuum interaction
as it ignores the frequency shift.
Expanding also the real frequency as $\omega_r=\omega_0 + \omega_1$ with $\omega_1\ll\omega_0$ yields, 
to second-order in both $\omega_1$ and $\gamma$, the components 
\begin{align}
  \text{Re}[\chi] =& \chi_{\text{adi}} + \omega_1\chi_{\text{adi}}^\prime - \gamma\chi_{\text{res}}^\prime, \label{eq:real_part}\\
  \text{Im}[\chi] =& \chi_{\text{res}} + \omega_1\chi_{\text{res}}^\prime + \gamma\chi_{\text{adi}}^\prime. \label{eq:imag_part}
\end{align}
Solving for $\gamma$, $\omega_1$ with $\chi=0$ in the vicinity of an embedded eigenvalue $\omega_0$ satisfying $\chi_{\text{adi}}(\omega_0) = 0$ gives
\begin{align}
  \gamma &= -\cos^2\theta \cdot \chi_{\text{res}}/\chi_{\text{adi}}^\prime,\label{eq:gamma_res} \\
  \omega_1 &= -\sin^2\theta \cdot \chi_{\text{res}}/\chi_{\text{res}}^\prime.\label{eq:omega_res}
\end{align}
where $\theta \equiv \arctan(\chi_{\text{res}}^\prime / \chi_{\text{adi}}^\prime)$ is the argument of the
complex derivative $(\chi^+)^\prime$ in the $(\chi_{\text{adi}}, \chi_{\text{res}})$ plane. %
This angle partitions the continuum interaction
between frequency and amplitude modulation of the discrete mode.
When $|\chi_{\text{res}}^\prime| \ll |\chi_{\text{adi}}^\prime|$,
all the interaction goes into amplitude modulation, producing Eq.~\ref{eq:continuum_damping}.
In the limit $|\chi_{\text{res}}^\prime| \gg |\chi_{\text{adi}}^\prime|$
the continuum resonance merely modulates the mode's frequency.

Equations~\ref{eq:gamma_res} and~\ref{eq:omega_res} are asymptotically valid,
while the weakly damped Z-pinch MHD instabilities have large growth rates and lie outside
their regime of validity.
However, the three possible consequences of continuum interaction, namely damping, excitation, 
or frequency modulation, persist qualitatively.

\subsection{Shear-driven instabilities}\label{subsec:shear_driven}
As seen in the previous section, a possible outcome of continuum coupling is instability.
When the continuum resonance is due to flow shear, such instability is loosely called Kelvin-Helmholtz.
It is necessary, though, to distinguish among the physical origins of shear-driven instability.
Classically, Kelvin-Helmholtz is an incompressible instability of a planar vortex sheet driven 
by the vortex-stretching mechanism.~\cite{Drazin_2004}
Many famous results on sheared-flow stability, such as Rayleigh's inflection point
criterion, describe only these incompressible dynamics of planar sheared flows.
Appendix~\ref{app:rayleigh} reviews the stability conditions of incompressible axisymmetric jets,
based upon which this paper considers the parabolic flow $v_{0z} = v_0(r/r_p)^2$ to avoid incompressible instability.

The shear-driven instabilities observed in this work instead originate from compressible dynamics 
owing to the trans-Alfv\'{e}nic sheared flow necessary to suppress the MHD kink instability.
These compressible instabilities arise from resonances of interior and exterior reflecting magnetosonic waves,
and are accordingly called reflection modes.~\cite{Cohn1983,Tam_1989}
The reflection modes are cylindrical analogues of the Mack modes occurring in transonic planar boundary layers.~\cite{Mack_1990,Schlichting2000}
The acoustics producing reflection modes are accurately described by WKB analysis, as detailed in Appendix~\ref{app:wkb},
except for one pernicious instability, the acoustic kink mode, which is discussed below.

\subsubsection{Three-family classification of marginal modes}\label{sec:three_family}
Three families of neutral modes underlie compressible shear-driven instability.
They arise from the radial structure of the linearized acoustic problem in sheared flow,
and are derived in Appendix~\ref{app:wkb} for a transonic current-free ($\vec{B}=0$) flow.
In trans-Alfv\'{e}nic sheared flow, compressibility shapes the dynamics, and the same three families
organize the MHD dispersion relations of Sec.~\ref{subsec:dispersion}.
The flow is assumed to increase in velocity going radially outwards,
with the opposite case merely rearranging the labeling of the families.
Assuming flow increasing in velocity outwards, the three families are:
\begin{enumerate}[label=(\roman*)]
  \item \emph{Interior modes} propagating in a
    subsonic cavity between an inner turning point
    (the axis for $m = 0$ or a potential barrier $r_{\min}$ for $|m| \geq 1$) 
    and an outer turning point $R_1$ where $\tilde{\omega}(R_1) = k c_s$.
    They are radially quantized, with each $n = 0, 1, 2, \ldots$
    satisfying a quantization condition (Eq.~\ref{eq:quantization_m}).
  \item \emph{Exterior modes} propagating between
    a turning point $R_2$ and the wall.
    When the wall is distant, the modes form a radiation continuum;
    when the wall is close, they quantize into discrete branches.
  \item \emph{Axis-localized modes} propagating with subsonic azimuthal and axial phase velocity 
    but not propagating radially, and hence existing only for $|m| \geq 1$.
    This family is quantized azimuthally by $m$, but per $m$ admits only a single
    radial mode localized near the axis with a pressure perturbation $P \propto I_m(\kappa r)$.
\end{enumerate}

\subsubsection{Summary of shear-driven instability modes}
Pairwise coupling between these families produces two instabilities: 
the \emph{reflection mode} from coupling of Families~(i) and~(ii),
and the \emph{interior kink and flute modes} from coupling of Families~(ii) and~(iii).
In both cases, the radiation of sound to the exterior, via Family (ii), 
accompanies the instability.
Together with the incompressible Kelvin-Helmholtz instability
(Appendix~\ref{app:rayleigh}), these correspond to the three families
of instability waves identified by Ref.~\citenum{Tam_1989}.
In each case the instability can be understood as mutual amplification
of positive- and negative-energy waves,
namely acoustic waves trapped between turning points for the compressible modes,
and propagating vorticity waves for the incompressible mode.~\cite{Cairns_1979, Hirota_2008, Hirota2009}

\subsubsection{Krein collisions and the Hamiltonian-Hopf bifurcation}\label{subsubsec:krein}
The mutual amplification of positive- and negative-energy waves is an instance of a general
phenomenon in the spectral theory of Hamiltonian systems, known there as the Krein collision,
or Hamiltonian-Hopf bifurcation.~\cite{Krein_1950, MacKay_1986, Kirillov_2013}
Although sheared flow renders the linearized force operator non-self-adjoint,~\cite{Frieman_1960}
the ideal system remains Hamiltonian, so every neutral mode carries a conserved signature,
namely the sign of its wave energy.
Krein's theorem states that an eigenvalue of definite signature cannot leave the real-frequency
axis under any parameter variation preserving the Hamiltonian structure; complex eigenvalues
emerge only where two neutral modes of opposite signature collide, merging and splitting into a
conjugate pair whose growing member is the instability.~\cite{Krein_1950, MacKay_1986}
The three neutral families of Sec.~\ref{sec:three_family} supply precisely these ingredients.
The interior and axis-localized families carry positive wave energy, while the exterior family,
Doppler-shifted by transonic flow, is a negative-energy wave in the sense of Ref.~\citenum{Cairns_1979}.
Each crossing of an exterior branch with an interior or axis branch in the $(k, \omega)$ plane is
therefore a Krein collision, and these opposite-signature crossings are where the instability
``bubbles'' appear along the marginal skeleton of Figs.~\ref{fig:current_free_m0}
and~\ref{fig:current_free_m1}, while same-signature crossings pass through without
destabilization.
The extension of Krein signature to the continuous spectrum, required to make the continuum-damping
competition of Sec.~\ref{subsubsec:continuum_damping} precise, is given in
Refs.~\citenum{Hirota_2008} and~\citenum{Hirota2009}.

\subsubsection{Reflection modes: coupling of Families (i) and (ii)}
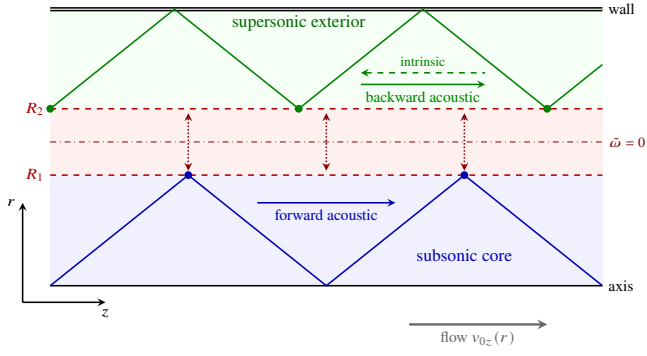
\begin{figure}[t]
\centering
\resizebox{\columnwidth}{!}{\begin{tikzpicture}[>=stealth]

  \def\zlen{10}        %
  \def\rmax{5.0}       %
  \def\Rone{2.0}       %
  \def\Rtwo{3.2}       %
  \def\rcrit{2.6}      %
  \def\raxis{0}        %

  \fill[blue!6] (0, \raxis) rectangle (\zlen, \Rone);
  \fill[red!6] (0, \Rone) rectangle (\zlen, \Rtwo);
  \fill[green!4] (0, \Rtwo) rectangle (\zlen, \rmax);

  \draw[red!70!black, thick, dashed] (0, \Rone) -- (\zlen, \Rone);
  \node[left, text=red!70!black, font=\footnotesize] at (0, \Rone) {$R_1$};

  \draw[red!70!black, thick, dashed] (0, \Rtwo) -- (\zlen, \Rtwo);
  \node[left, text=red!70!black, font=\footnotesize] at (0, \Rtwo) {$R_2$};

  \draw[red!50!black, thin, dash dot] (0, \rcrit) -- (\zlen, \rcrit);
  \node[right, text=red!50!black, font=\scriptsize] at (\zlen, \rcrit)
    {$\tilde{\omega} = 0$};

  \draw[thick] (0, \raxis) -- (\zlen, \raxis);
  \node[right, font=\footnotesize] at (\zlen, \raxis) {axis};

  \draw[thick, double] (0, \rmax) -- (\zlen, \rmax);
  \node[right, font=\footnotesize] at (\zlen, \rmax) {wall};

  \draw[blue!70!black, thick]
    (0, 0) -- (2.5, \Rone) -- (5, 0)
    -- (7.5, \Rone) -- (10, 0);

  \foreach \z in {2.5, 7.5} {
    \fill[blue!70!black] (\z, \Rone) circle (2pt);
  }

  \draw[green!50!black, thick]
    (0, \Rtwo) -- (2.25, \rmax) -- (4.5, \Rtwo)
    -- (6.75, \rmax) -- (9.0, \Rtwo) -- (10, 4.0);

  \foreach \z in {0, 4.5, 9.0} {
    \fill[green!50!black] (\z, \Rtwo) circle (2pt);
  }

  \foreach \z in {2.5, 5, 7.5} {
    \draw[red!50!black, thick, densely dotted, <->]
      (\z, \Rone+0.08) -- (\z, \Rtwo-0.08);
  }

  \draw[black!60, very thick, ->] (6.5, -0.7) -- (9.0, -0.7)
    node[midway, below, font=\footnotesize] {flow $v_{0z}(r)$};

  \draw[blue!70!black, thick, ->] (3.75, 1.5) -- (6.25, 1.5)
    node[midway, below, font=\footnotesize] {forward acoustic};
  \draw[green!50!black, thick, ->] (5.625, 3.64) -- (7.875, 3.64)
    node[midway, below, font=\footnotesize] {backward acoustic};
  \draw[green!50!black, thick, <-, dashed] (5.625, 3.85) -- (7.875, 3.85)
    node[midway, above, font=\scriptsize] {intrinsic};

  \node[blue!70!black, font=\small] at (7.5, 0.55)
    {subsonic core};
  \node[green!40!black, font=\small] at (4.5, 4.75)
    {supersonic exterior};

  \draw[thick, ->] (-0.5, -0.3) -- (-0.5, 1.5)
    node[left, font=\footnotesize] {$r$};
  \draw[thick, ->] (-0.5, -0.3) -- (1.0, -0.3)
    node[below, font=\footnotesize] {$z$};

\end{tikzpicture}}
\caption{Schematic of the reflection mode mechanism in the $(r,z)$ plane.
Upstream-propagating waves are trapped in the subsonic core ($k_r^2 > 0$),
reflecting between the axis and the sonic turning point $\tilde{\omega} = c_s k$.
Beyond this boundary the radial wavenumber becomes imaginary ($k_r^2 < 0$)
and the wave amplitude decays evanescently, with outward energy leakage
that drives the instability.}
\label{fig:reflection_mode}
\end{figure}

Reflection modes, present for all $m$,
arise from the coupling of Family~(i) with Family (ii).
Sound waves propagating co-flow in the subsonic core reflect radially
at a sonic turning point. 
The mode resonates with counter-flow propagating, co-flow advected, exterior modes 
themselves reflected off a sonic turning point, shown schematically in Fig.~\ref{fig:reflection_mode}.

\begin{figure}[htbp]
\centering
\makebox[\columnwidth][l]{(a)}\\
\includegraphics[width=\columnwidth]{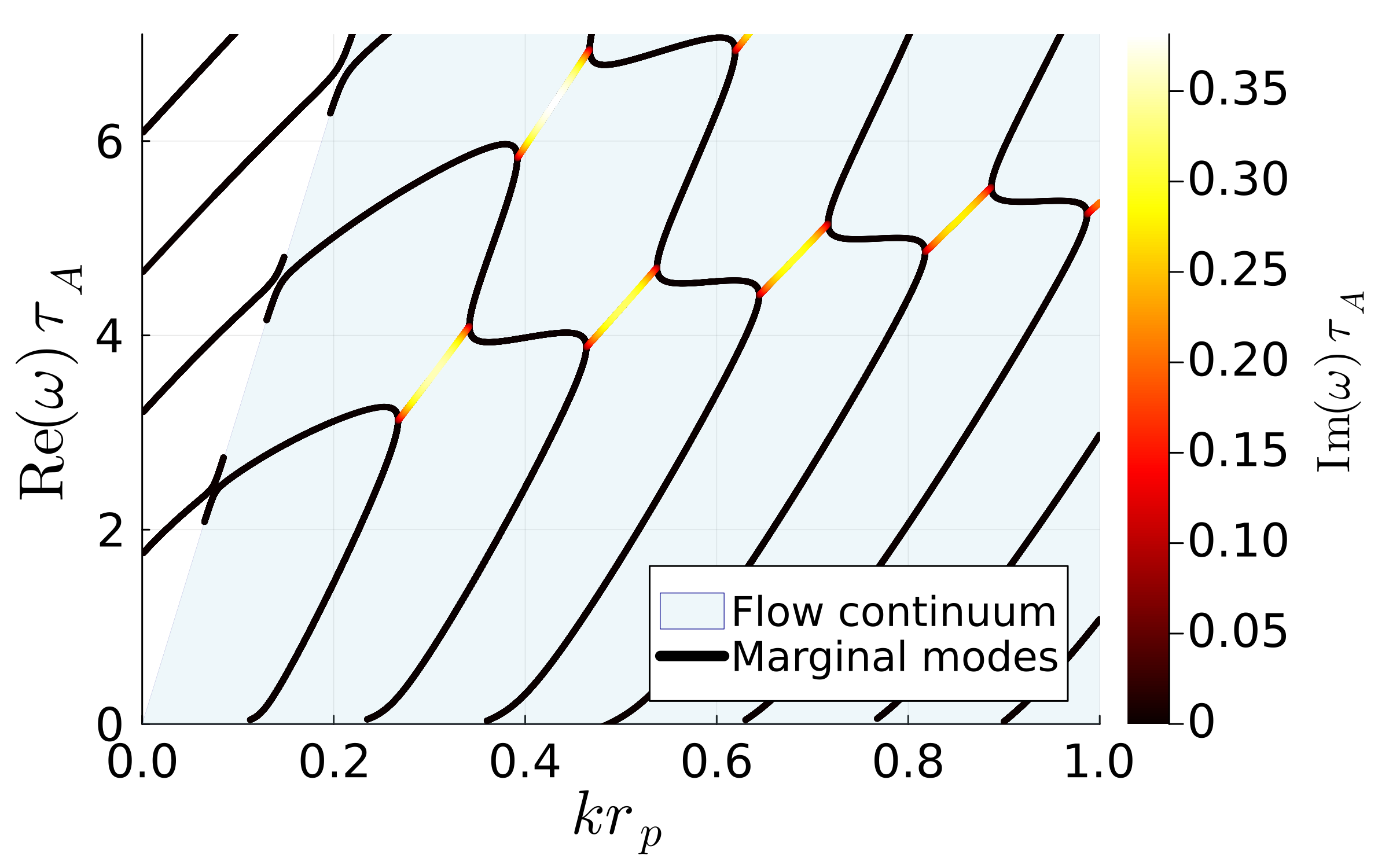}\\[4pt]
\makebox[\columnwidth][l]{(b)}\\
\includegraphics[width=\columnwidth]{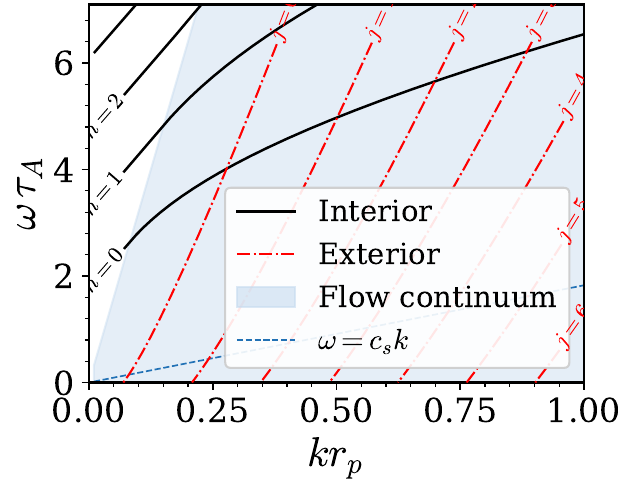}
\caption{(a)~Dispersion relation for $m=0$ modes of a current-free ($B_\theta=0$)
parabolic shear flow $v_{0z} = v_0(r/r_p)^2$ with $v_0 = 2\,v_A^*$
($M_s = v_0/c_s \approx 1.1$) and wall at $r_w = 4r_p$.
The shaded region is the sonic continuum $\omega = kv_{0z}(r)$;
colored segments indicate unstable modes with $\mathrm{Im}(\omega) > 0$.
Marginal modes in the continuum are embedded eigenvalues computed by 
$\text{Re}(\chi^+)=0$ (cf.~Section~\ref{subsubsec:embedded}).
(b)~WKB dispersion branches for the same configuration.
The black curves are interior-cavity modes of Family (i),
anchored at the exact $k=0$ eigenfrequencies (zeros of $J_1$).
The red dashed curves are exterior-cavity modes of Family (ii).
Reflection modes occur at intersections of the two families.}
\label{fig:current_free_m0}
\end{figure}

When the wall is close enough to quantize Family~(ii), the
two families cross and resonate at specific $(k, \omega)$ pairs
(Figs.~\ref{fig:current_free_m0}b and~\ref{fig:current_free_m1}b)
generating instability.
When the wall is distant, Family~(ii) merges into a continuum and 
Family~(i) is destabilized on a continuous frequency band.
While $r_w=4r_p$ is distant enough for the MHD static kink mode to be 
negligibly impacted by a conducting wall, the exterior acoustic modes 
are not close to a continuum until $r_w \gtrsim 10 r_p$.

\subsubsection{Acoustic kink instability: Families (ii) and (iii)}
\begin{figure}[htbp]
\centering
\makebox[\columnwidth][l]{(a)}\\
\includegraphics[width=\columnwidth]{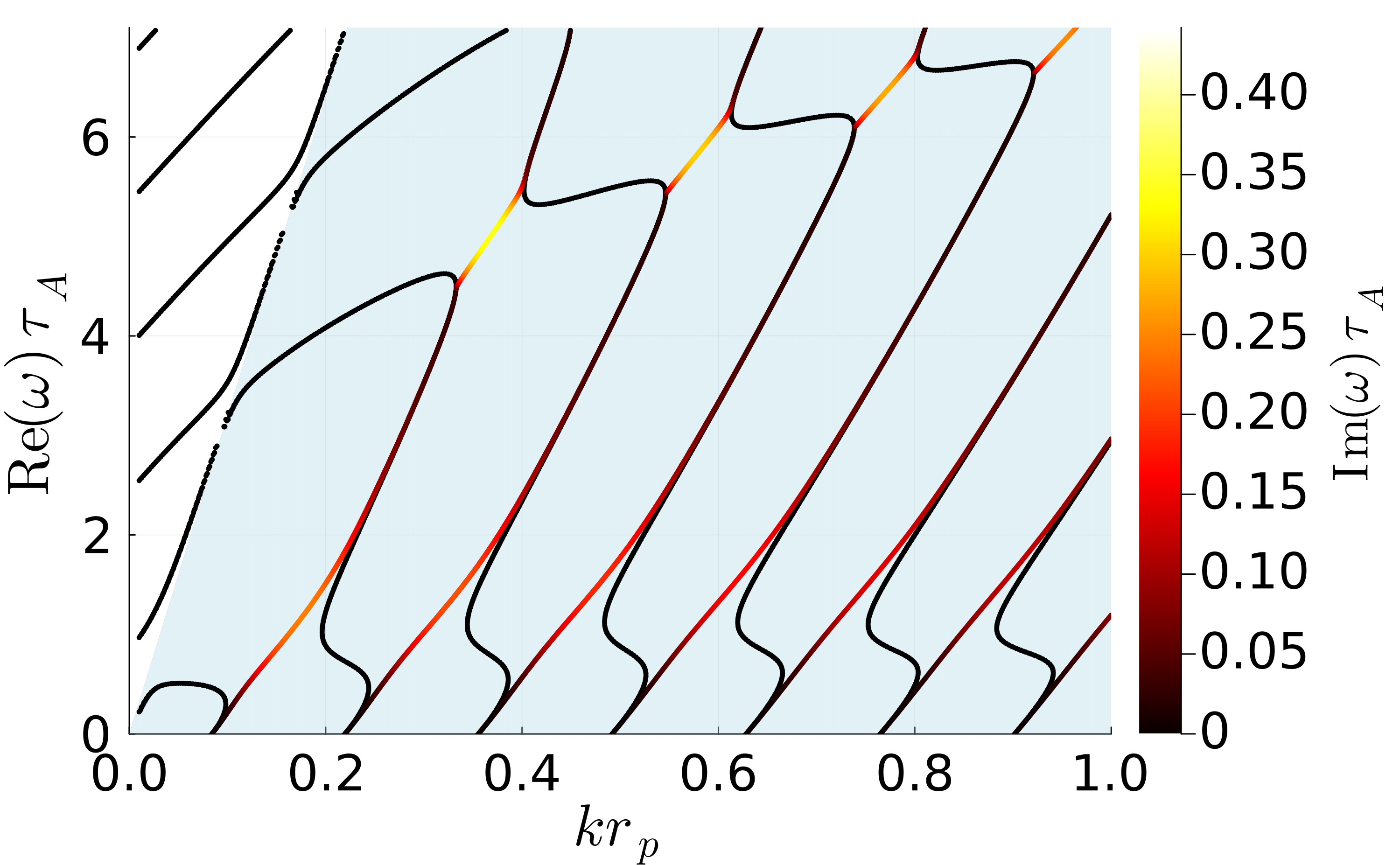}\\[4pt]
\makebox[\columnwidth][l]{(b)}\\
\includegraphics[width=\columnwidth]{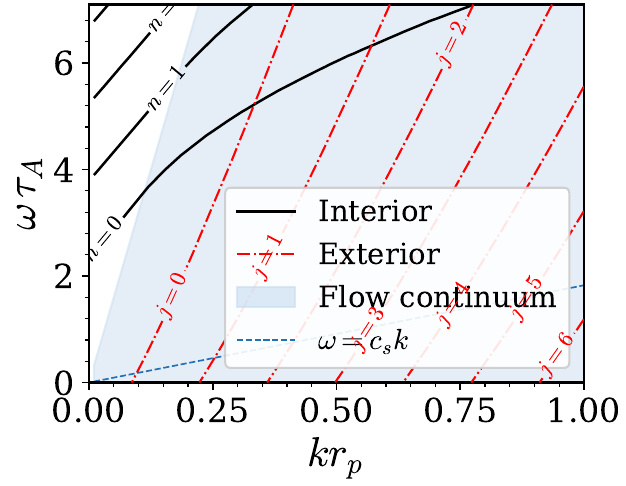}
\caption{(a)~Dispersion relation for $m=1$ modes of the
same flow as Fig.~\ref{fig:current_free_m0}
($v_0 = 2\,v_A^*$, $r_w = 4r_p$).
All three families are visible: interior acoustic modes (Family~i),
exterior acoustic modes (Family~ii), and axis-localized mode (Family~iii) around $\omega\approx\sqrt{2}\,v_A^*/r_p$.
(b)~WKB dispersion branches for the same configuration,
with black curves for Family (i), anchored at the exact $k=0$
eigenfrequencies (zeros of $J_1'$), and red dash-dot curves for Family (ii).
The WKB analysis captures only Families~(i) and~(ii).
Family~(iii), responsible for the kink mode in (a),
requires the full computation.}
\label{fig:current_free_m1}
\end{figure}

At the crossings in the $(k, \omega)$ plane where
Families~(ii) and~(iii) resonate (Fig.~\ref{fig:current_free_m1}a),
the axis localized mode couples to the exterior acoustic field.
For $m=1$, this coupling converts the neutral axis sound wave into a kink instability.

Having described the physics of the shear-driven instabilities,
the following section turns to a numerical investigation of the complete problem,
namely the shear-flow stabilization of MHD instabilities and the simultaneous 
development of shear-driven instability in the inviscid/irresistive limit.

\section{Computational results}
\label{sec:results}

The spectral theory of Section~\ref{sec:spectrum}
identified two competing effects of sheared flow, namely
continuum damping of discrete MHD instabilities
and excitation of compressible instabilities.
This section presents numerical solutions of the
eigenvalue problem that quantify these effects
for the Z-pinch Bennett equilibrium with parabolic sheared flow.
Section~\ref{subsec:setup} specifies the normalization, equilibrium profiles,
and boundary conditions,
Sec.~\ref{subsec:method} discusses the numerical method,
Sec.~\ref{subsec:m0_stabilization} demonstrates sub-Alfv\'{e}nic stabilization
of $m=0$ interchange close to the Kadomtsev-marginal boundary,
and Sec.~\ref{subsec:dispersion} discusses shear-flow stabilization of the $m=1$ kink mode 
by presenting the full dispersion relations at trans-Alfv\'{e}nic shear
where compressible instabilities emerge.
Growth rates as functions of wavenumber and flow amplitude
have been reported elsewhere;~\cite{Angus_2020}
the present study focuses instead on the dispersion relations
that underlie them,
revealing the spectral mechanisms responsible for stabilization
and the emergence of new instabilities.

\subsection{Problem set up and boundary conditions}\label{subsec:setup}

All results are presented with lengths normalized to the pinch radius $r_p$,
velocities to the reference Alfv\'{e}n speed
$v_A^* = B_{\theta,\mathrm{max}}/\sqrt{\mu_0\rho_0}$,
and times to the Alfv\'{e}n time $\tau_A \equiv r_p/v_A^*$,
where $B_{\theta,\mathrm{max}}$ is the peak azimuthal field
and $\rho_0$ the on-axis density of the Bennett equilibrium.
This reference speed is held fixed and used throughout,
including the current-free comparisons of Sec.~\ref{subsec:dispersion}
and Appendix~\ref{app:wkb}, whose field-free flow carries no Alfv\'{e}n speed of its own.
For the isothermal Bennett equilibrium 
$v_A^* = v_{ti}/\sqrt{2}$ with $v_{ti} = \sqrt{kT_i/m_i}$.
The specific heat ratio is taken to be $\gamma=5/3$.
The computational domain is $r\in[0, r_w]$ with $r_w/r_p = 4$.
The boundary-value problem, Eq.~\ref{eq:ode_sys}, requires regularity at the axis:
\begin{equation}\label{eq:axis_bc}
  (r\xi_r)\big|_{r=0} = 0, \qquad
  P\big|_{r=0} = \begin{cases} \text{finite} & m=0 \\ 0 & |m|\geq 1 \end{cases},
\end{equation}
derived from Frobenius analysis of Eq.~\ref{eq:ode_sys} near $r=0$,~\cite{Appl1992,Angus_2020}
which determines the leading-order behavior of $(r\xi_r, P)$ and thereby
selects the one-dimensional subspace $\vec{u}(0)$ appearing in Eq.~\ref{eq:dispersion_function}.
The wall at $r=r_w$ imposes zero displacement,
\begin{equation}\label{eq:wall_bc}
  (r\xi_r)\big|_{r=r_w} = 0.
\end{equation}

Three equilibrium profiles are considered:
The polytropic equilibrium of Eqs.~\ref{eq:poly1}--\ref{eq:poly2} with $\alpha=2$ (Bennett) to study
$m=1$ stability, with varying $\alpha$ to study $m=0$ stability, and
a ``current-free'' configuration with uniform pressure and no azimuthal field
($p_0=1$, $B_{0\theta}=0$, $\rho_0=1$) to isolate the shear-driven instabilities from MHD modes entirely.

A parabolic shear flow $v_{0z}(r) = v_0\,(r/r_p)^2$ is chosen,
where $v_0 = v_{0z}(r_p)$ is the flow speed at the pinch radius.
This profile lacks a generalized inflection point (Appendix~\ref{app:rayleigh})
and is therefore stable to the incompressible Kelvin-Helmholtz instability (at least with zero magnetic field),
making it the preferred choice for studying shear-stabilization of ideal MHD modes
and the shear-driven modes arising at trans-Alfv\'{e}nic levels of shear.

For the isothermal Bennett equilibrium ($\alpha = 2$),
the characteristic speeds at the pinch radius are
$c_s = \sqrt{2\gamma}\,v_A^* \approx 1.83\,v_A^*$
and $v_a(r_p) = 2\,v_A^*$,
so the sound speed and Alfv\'{e}n speed are comparable.
The local Alfv\'{e}n speed exceeds the reference $v_A^*$
because the on-axis density normalizing $v_A^*$
is four times the local density at the pinch radius.
The two flow amplitudes studied,
$v_0 = 2\,v_A^*$ and $v_0 = 4\,v_A^*$,
correspond to local Alfv\'{e}n Mach numbers
$M_A(r_p) = v_0/v_a(r_p) = 1$ and $2$ at the pinch radius,
and sonic Mach numbers $M_s(r_p) = v_0/c_s \approx 1.1$ and $2.2$.
The first case is marginally trans-Alfv\'{e}nic,
and the second is strongly trans-Alfv\'{e}nic,
placing the system well into the regime
where the Alfv\'{e}n spectral ``gap'' has closed.

\subsection{Method of solution}\label{subsec:method}

The dispersion function $\chi(\omega, k)$, defined in Eq.~\ref{eq:dispersion_function},
is computed by numerical integration of the ODE system, Eq.~\ref{eq:ode_sys}, from axis to wall.
Unlike finite-difference or Galerkin discretizations,
which reduce the boundary-value problem to a matrix eigenvalue problem
and return only converged eigenvalues,
direct integration constructs $\chi$ as a pointwise-evaluable function
of $(\omega, k)$ whose analytic structure is preserved at every point in parameter space, 
in particular the Plemelj decomposition into adiabatic and resonant parts
(Sec.~\ref{sec:marginal}).
Further, integration can adapt to resolve the steep gradients
developing near the continuous spectrum, where eigenfunctions become highly localized.
The principal advantage of direct integration over grid-based discretizations
is that the singular structure of the ODE at the continuous spectrum
may be treated by complex-analytic techniques rather than absorbed into numerical diffusion.

The above construction is closely related to the Spectral Web technique for spectral analysis of
steady MHD equilibria.~\cite{Goedbloed_2018a,Goedbloed_2018b}
Although equilibria with sheared flows are non-self-adjoint, the Spectral Web takes advantage of the
self-adjoint component force and Doppler-Coriolis forms and their alternator, resulting in a pair of 
solution and conjugate paths in the complex-frequency plane whose crossings indicate the eigenvalues.
Mode counting in the dispersion function construction $\chi(\omega,k)$ arises from application of the 
Nyquist method; in the Spectral Web, mode counting arises from the oscillation structure of the 
component self-adjoint forms.

Now, in the present work's dispersion function method, unstable modes are searched for as follows.
First, the adiabatic dispersion function $\chi^+(\omega_r, k)$,
regularized at $\varepsilon = 10^{-5}$,
is evaluated on a grid in $(\omega_r, k)$ and its zero-contours are extracted.
These contours are the marginal stability boundaries
where discrete modes emerge from, or are absorbed into, the continuous spectrum.
Between adjacent zero-contours lie gaps in which unstable branches reside.

In the second stage, a trial eigenvalue is seeded within each gap where 
instability is expected, and the solution followed by pseudo-arclength continuation
implemented via BifurcationKit.jl.~\cite{Veltz2020BifurcationKit}
This continuation approach exploits the implicit function theorem
to trace complete dispersion branches through turning points and bifurcations,
providing systematic coverage of the parameter space.

\subsection{Sub-Alfv\'{e}nic shear stabilization of interchange instability}\label{subsec:m0_stabilization}
\begin{figure*}[htbp]
\centering
\begin{subfigure}[b]{0.48\textwidth}
    \includegraphics[width=\textwidth]{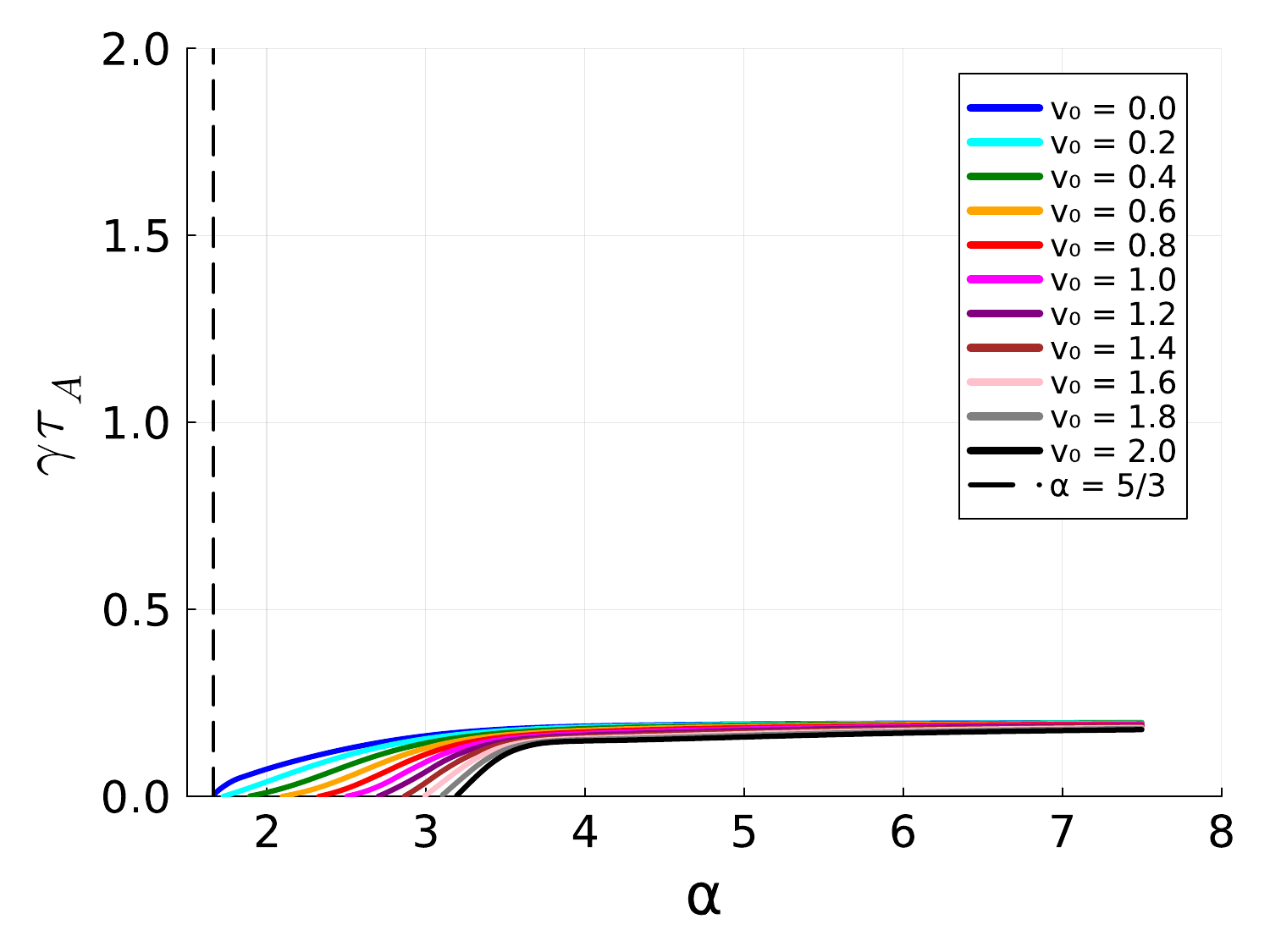}
    \caption{$kr_p = 0.1$}
    \label{fig:m0_shear_a}
\end{subfigure}
\hfill
\begin{subfigure}[b]{0.48\textwidth}
    \includegraphics[width=\textwidth]{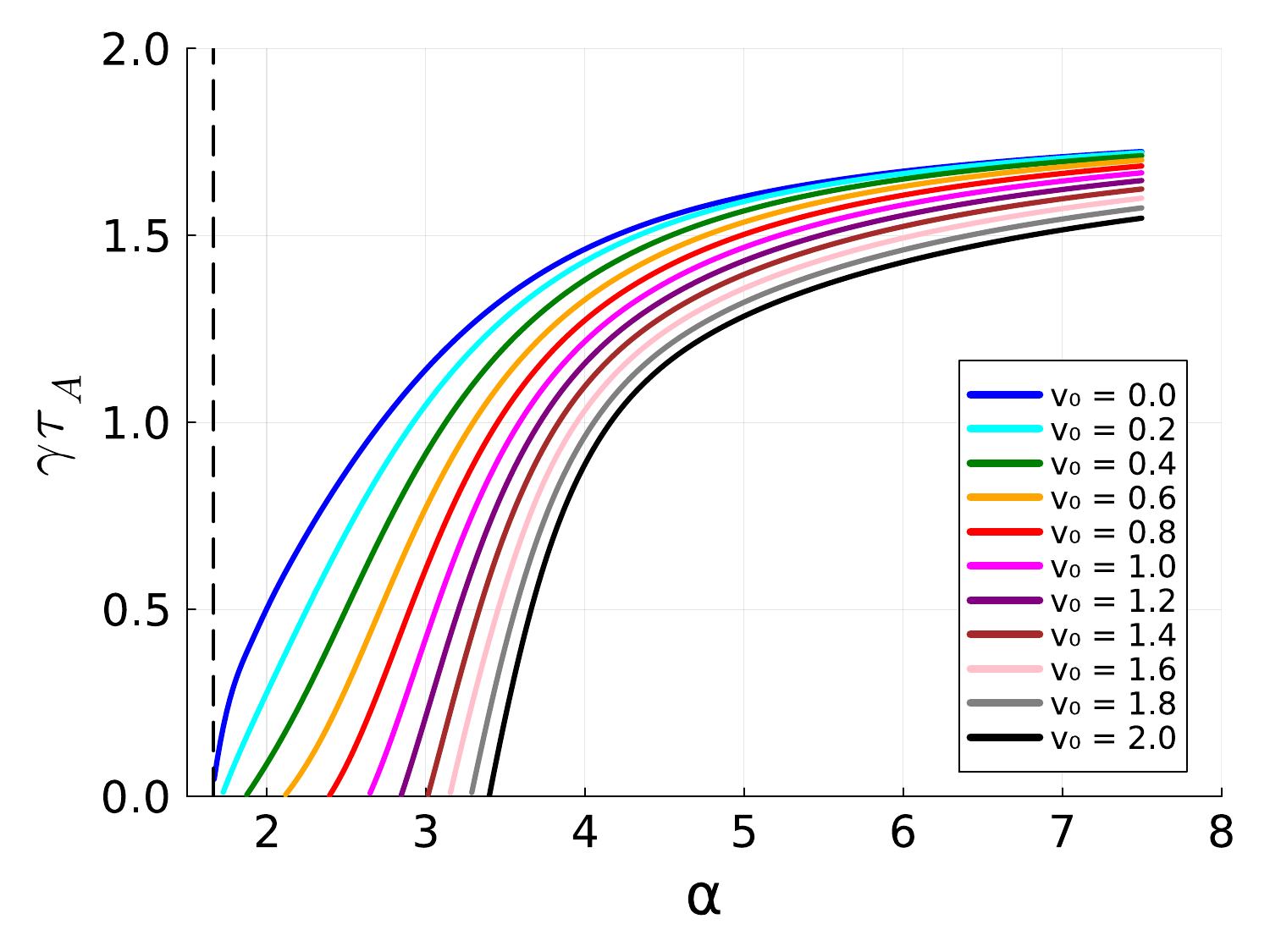}
    \caption{$kr_p = 1.0$}
    \label{fig:m0_shear_b}
\end{subfigure}
\caption{Flow stabilization of $m=0$ interchange modes
for the polytropic equilibrium of Eqs.~\ref{eq:poly1}--\ref{eq:poly2}
with parabolic sheared flow $v_{0z} = v_0(r/r_p)^2$ and wall at $r_w = 4r_p$,
illustrated by plotting growth rate as a function of $\alpha$.
Legend labels denote the magnitude of the flow shear;
the speed at the pinch radius is $v_{0z}(r_p) = v_0$,
so $v_0 = 2$ corresponds to $M_A(r_p) = 1$
since the local Alfv\'{e}n speed is $v_a(r_p) = 2\,v_A^*$.
Sub-Alfv\'{e}nic shear flows are seen to stabilize
profiles well beyond the static Kadomtsev condition $\alpha \leq \gamma$
(cf.\ Fig.~\ref{fig:static_growth_vs_alpha}).
At large $\alpha$, even strong shear does little, showing that strongly super-marginal profiles
resist shear-flow stabilization.}
\label{fig:m0_shear_stabilization}
\end{figure*}

The first study considers interchange stability with sub- to trans-Alfv\'{e}nic shear flows.
Figure~\ref{fig:m0_shear_stabilization} shows the $m=0$ growth rate
across the $(\alpha, v_0)$ plane for the polytropic equilibrium at long and short wavelength.
The key result is that sub-Alfv\'{e}nic shear flows ($v_0 \lesssim v_a(r_p)$, i.e.\ $M_A \lesssim 1$)
stabilize profiles well beyond the static Kadomtsev limit $\alpha \leq \gamma$.

The mechanism is continuum damping induced by radially sheared axial flow.
For $m=0$, resonance occurs at simply $\omega(r) = kv_z(r)$ 
(Sec.~\ref{subsec:MHD_continuum}),
the sheared flow continuum overlaps with the unstable interchange 
eigenfrequencies, and the resonant interaction damps the instability
(Sec.~\ref{subsubsec:continuum_damping}).
The stabilization is monotonic in $v_0$, with stronger shear
producing a broader continuum overlap and more effective damping.

At shorter wavelength (Fig.~\ref{fig:m0_shear_b}, $kr_p = 1.0$),
the static growth rate is large
(Fig.~\ref{fig:static_growth_vs_alpha})
but the continuum overlap is also broad,
and the sheared flow extends the stability boundary
to $\alpha \gg \gamma$.
At long wavelength (Fig.~\ref{fig:m0_shear_a}, $kr_p = 0.1$),
the interchange is more resistant to stabilization.
Although the static growth rate is smaller,
the narrower continuum overlap at small $k$
makes the damping less effective,
and stronger flows are needed to achieve
the same relative suppression.
This result is consistent with experimental observations
and two-fluid modeling of sub-Alfv\'{e}nic shear flows stabilizing
interchange modes in sheared-flow Z~pinches.~\cite{Shumlak_2020, Meier_2021}

Sub-Alfv\'{e}nic sheared flows are sufficient to stabilize near-marginal profiles, such
as the Bennett equilibrium.
The $m=1$ kink, by contrast, resists sub-Alfv\'{e}nic stabilization
as the shear Alfv\'{e}n and slow magnetosonic resonances
shield the discrete kink eigenvalue from the continuum (Sec.~\ref{subsec:MHD_continuum}),
and trans-Alfv\'{e}nic flows are required to close the gap
and access the continuum damping mechanism.
The consequences of this asymmetry are explored in the following section.

\subsection{Dispersion relations at trans-Alfv\'{e}nic shear}\label{subsec:dispersion}
The preceding section showed that sub-Alfv\'{e}nic sheared flow
stabilizes $m=0$ modes provided the equilibrium is not greatly super-marginal.
This section now examines the full dispersion relations above the Alfv\'{e}nic threshold,
where the kink mode is damped by flow shear and qualitatively new instabilities appear.
Figures~\ref{fig:m0_dispersion} and~\ref{fig:m1_dispersion}
show the Bennett profile ($\alpha=2$) at $v_0 = 2\,v_A^*$ ($M_A = 1$)
and $v_0 = 4\,v_A^*$ ($M_A = 2$).
The zero-contours of $\chi^+(\omega_r, k)$ (black curves)
provide the skeleton of marginal stability boundaries;
the MHD continuum bands are shaded for $m=1$
(Alfv\'{e}n in blue, magnetosonic in red, where applicable).
Unstable branches (colored by growth rate) are traced by continuation
from seeds placed within gaps in this skeleton.

\subsubsection{Axisymmetric modes}
At both $v_0 = 2\,v_A^*$ ($M_A = 1$) and $4\,v_A^*$ ($M_A = 2$),
the interchange instability is entirely eliminated at all wavenumbers
with no such branch observed in Figures~\ref{fig:m0_v2} and~\ref{fig:m0_v4}.
Because $m=0$ modes have no Alfv\'{e}n spectral gap
and the Bennett equilibrium is already close to the interchange stability threshold,
even modest flows provide sufficient continuum overlap to damp the interchange completely.

The instabilities that remain are compressible shear-driven modes,
identifiable by comparison with the current-free dispersion relation
of Fig.~\ref{fig:current_free_m0}.
At $M_A = 2$, the stronger shear opens more spectral gaps
and correspondingly more reflection branches appear.
In the trans-Alfv\'{e}nic regime, the $m=0$ stability is broadly similar, phenomenologically, 
to the current-free reflection modes, provided the equilibrium profile is not excessively far from marginal.
The growth rates are comparable to the sonic crossing time because the modes
arise from magnetosonic wave emission, which corresponds to the Alfv\'{e}n time 
due to the unity-$\beta$ nature of the Z pinch.
Such modes would appear in experiment as Kelvin-Helmholtz-like waves in the shear layer
accompanied by radiation of magnetosonic waves to the exterior.

\subsubsection{Non-axisymmetric modes}

\begin{figure}[htbp]
\centering
\includegraphics[width=\columnwidth]{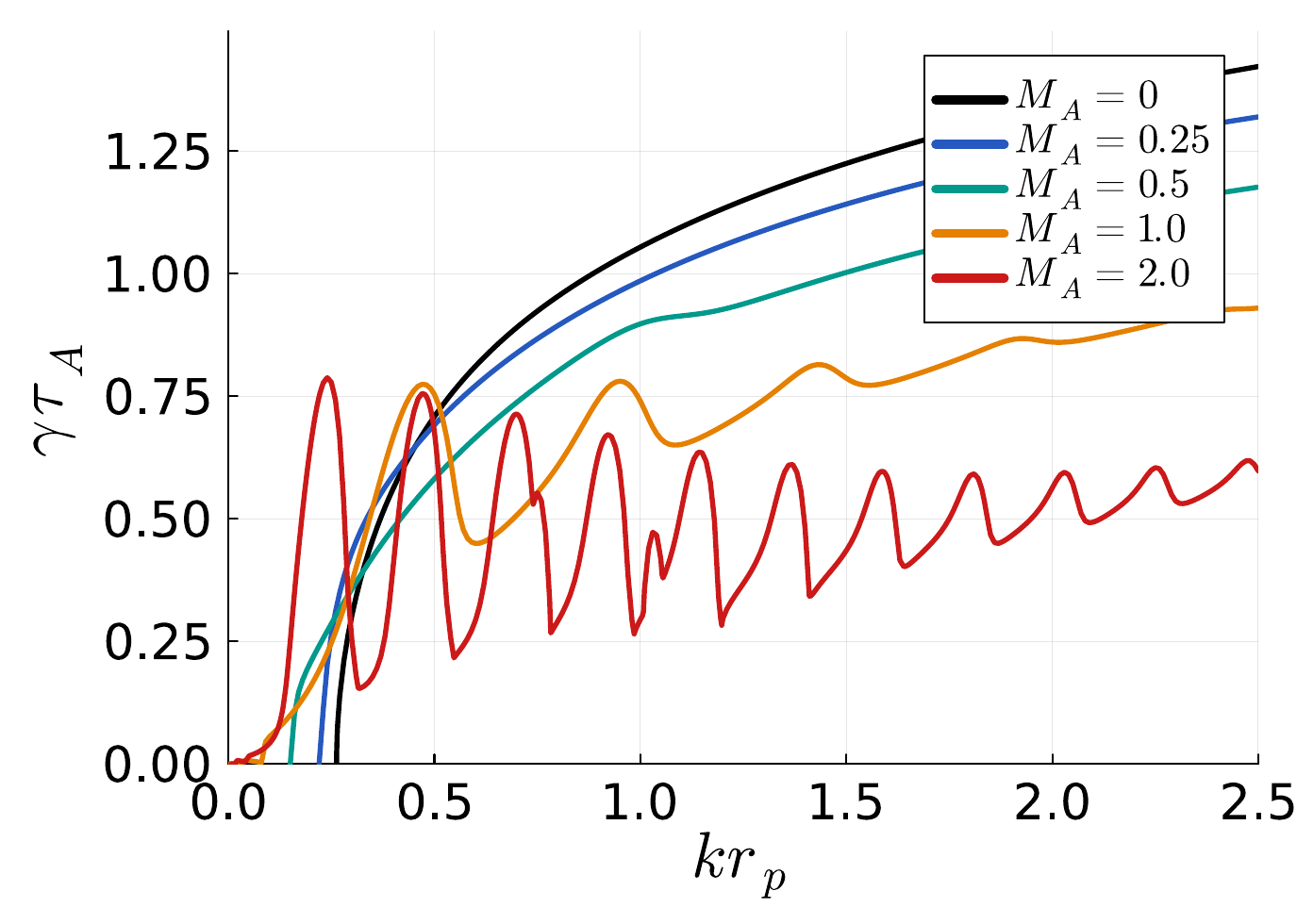}
\caption{Growth rate $\gamma\tau_A$ at $m=1$
for the Bennett equilibrium with wall at $r_w = 4r_p$ and flow $v_{0z}=v_0(r/r_p)^2$
at Alfv\'{e}nic Mach numbers $M_A = v_0/(2v_A^*)$.
Sub-Alfv\'{e}nic shear ($M_A \le 0.5$) reduces the growth rate only weakly,
and at $M_A = 1$ the kink is a hybrid MHD-acoustic branch of diminished growth.
The maximum growth rate at each $kr_p$ is plotted for $M_A=2$, because the shear-driven acoustic kink
exhibits wall-induced resonance patterns and multiple solutions, producing a jagged curve.
With $r_w\gg r_p$, the acoustic kink would be smoothly destabilized as the axis magnetosonic wave couples
to an exterior magnetosonic continuum.}
\label{fig:kink_growth_vs_k}
\end{figure}

The $m=1$ modes are fundamentally different from the $m=0$ modes because
the ``Alfv\'{e}n gap'' in the continuum shields the MHD kink from continuum damping,
so that trans-Alfv\'{e}nic flow is required merely to access the stabilization mechanism.
In doing so, this fast sheared flow simultaneously activates compressible instabilities which are completely 
undamped in the ideal model.
This is apparent from Fig.~\ref{fig:kink_growth_vs_k} which summarizes the growth rate of the $m=1$ kink
across the sub- to trans-Alfv\'{e}nic transition.
The growth rates exhibit resonance peaks, which are explained by examining the underlying dispersion relations
plotted in Figs.~\ref{fig:m1_v2} and~\ref{fig:m1_v4}.
At $M_A = 1$, the kink is partially stabilized at long wavelength ($kr_p \lesssim 1$),
where the Doppler-shifted Alfv\'{e}n and magnetosonic continuum boundaries
intersect the kink branch and continuum damping acts.
Within the spectral gaps between these intersections, however, the kink remains unstable.
At short wavelength ($kr_p \gtrsim 1$), the kink is Doppler-shifted to positive phase velocity
but retains near-zero group velocity.
This is characteristic of the acoustic kink discussed in Sec.~\ref{subsec:surface_mode},
which requires sub-acoustic phase velocity to propagate along the axis.
The trans-Alfv\'{e}nic flow required to close the gap
and access continuum damping at long wavelength
simultaneously exposes the system to acoustic kink instability
via magnetosonic radiation, broadly similar in the current-free case (Fig.~\ref{fig:current_free_m1}).
Therefore, the main instability branch with $M_A=1$ should be understood as a hybrid of the original MHD kink 
and the acoustic kink instabilities.

The $m=1$ spectrum also hosts unstable reflection modes, 
like in the current-free problem (Sec.~\ref{sec:three_family}).
In the MHD case, however, the discrete mode within these magnetosonic 
gaps can interact with the Alfv\'{e}n continuum.
The resulting continuum damping can suppress
$m=1$ reflection mode instabilities entirely when they would otherwise be only weakly destabilized.
A count of unstable eigenvalues in the upper spectral gap of
Fig.~\ref{fig:m1_v2} via the Nyquist method confirms that none exist there.
Specifically, at fixed wavenumber, the winding number of $\chi(\omega; k)$ around a closed contour
enclosing the gap in the upper-half complex-$\omega$ plane is zero,
which by the argument principle equals the number of complex eigenvalues
with positive growth rate enclosed.
The reflection-mode instability that would otherwise occupy this gap
is fully suppressed by Alfv\'{e}n continuum damping at $M_A=1$;
instability appears only at greater $v_0$.

Figure~\ref{fig:m1_v2_zoom} zooms into the region near the Alfv\'{e}n frequency
in the $M_A = 1$ dispersion relation.
The accumulation modes near the Alfv\'{e}n frequency
are seen to be destabilized by coupling to exterior magnetosonic waves.
These tertiary instabilities are distinct from both the primary kink
and the reflection modes discussed above.
Their growth rates are small compared to the primary kink
and reflection mode instabilities.

The flow is strongly super-Alfv\'{e}nic at $M_A = 2$ (Fig.~\ref{fig:m1_v4})
and the Alfv\'{e}n gap protecting the kink mode fully closes at long wavelength.
The incompressible MHD kink is absorbed into the continuum
and is replaced by compressible modes.
The axis mode (Family~iii of Sec.~\ref{sec:three_family}) and exterior magnetosonic branches
couple and form an acoustic kink instability
analogous to the current-free case described in Appendix~\ref{app:wkb}.
Compared with the current-free case,
the MHD dispersion relation exhibits fewer exterior branches,
because the fast magnetosonic speed $c_f = \sqrt{c_s^2 + v_a^2}$
exceeds the sound speed and shrinks the exterior acoustic cavity.
Its convergence to an external continuum requires the wall to be much further away than $r_w=4r_p$.
The acoustic kink growth rates are smaller than the static kink
but persist across a broad range of wavenumbers,
and their ultimate fate in the presence of dissipation
remains an open question with implications
for sheared-flow Z-pinch stability.
This question is outside of the realm of ideal MHD.

\begin{figure*}[htbp]
\centering
\begin{subfigure}[b]{0.48\textwidth}
    \includegraphics[width=\textwidth]{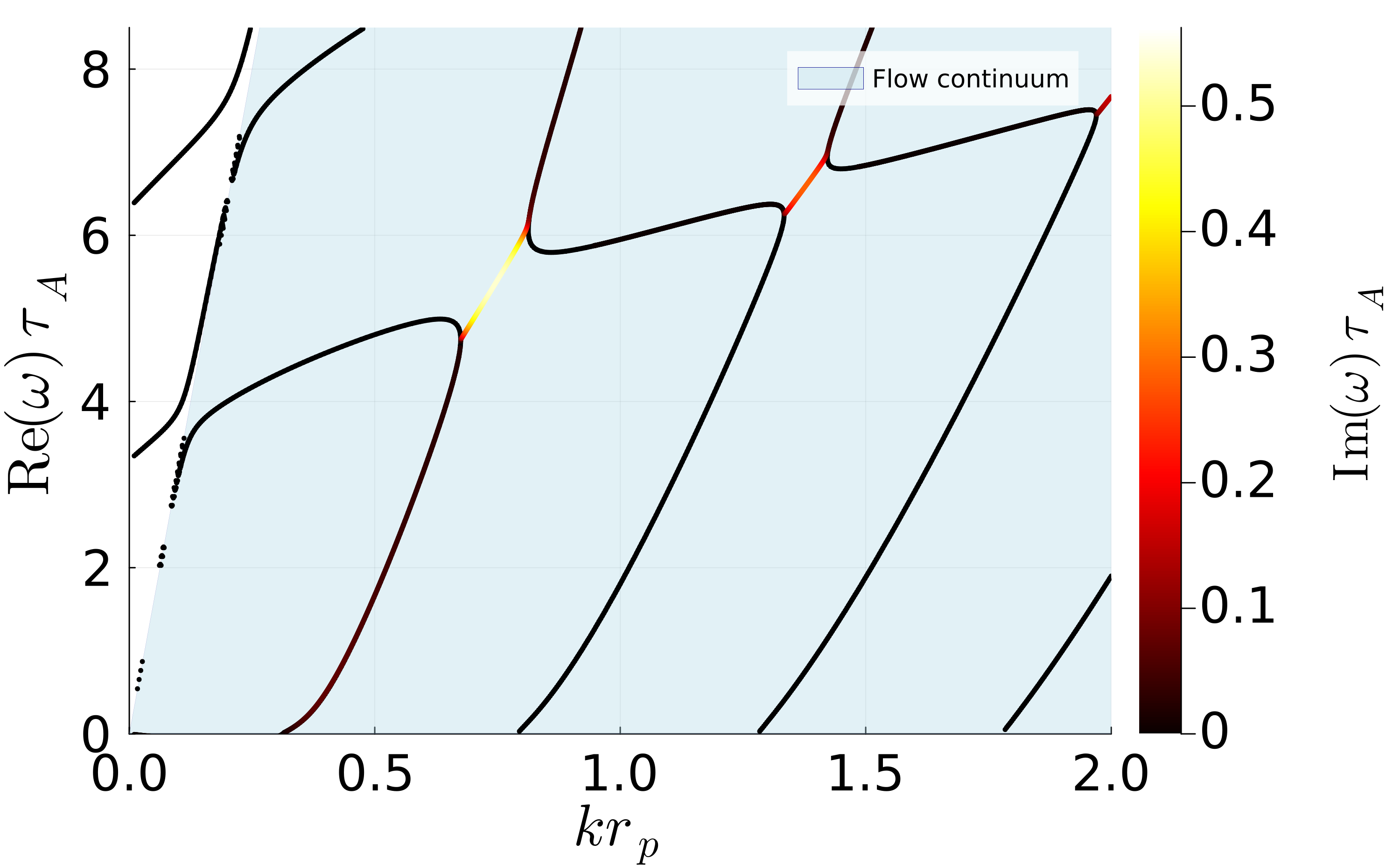}
    \caption{$M_A = 1$ ($v_0 = 2\,v_A^*$)}
    \label{fig:m0_v2}
\end{subfigure}
\hfill
\begin{subfigure}[b]{0.48\textwidth}
    \includegraphics[width=\textwidth]{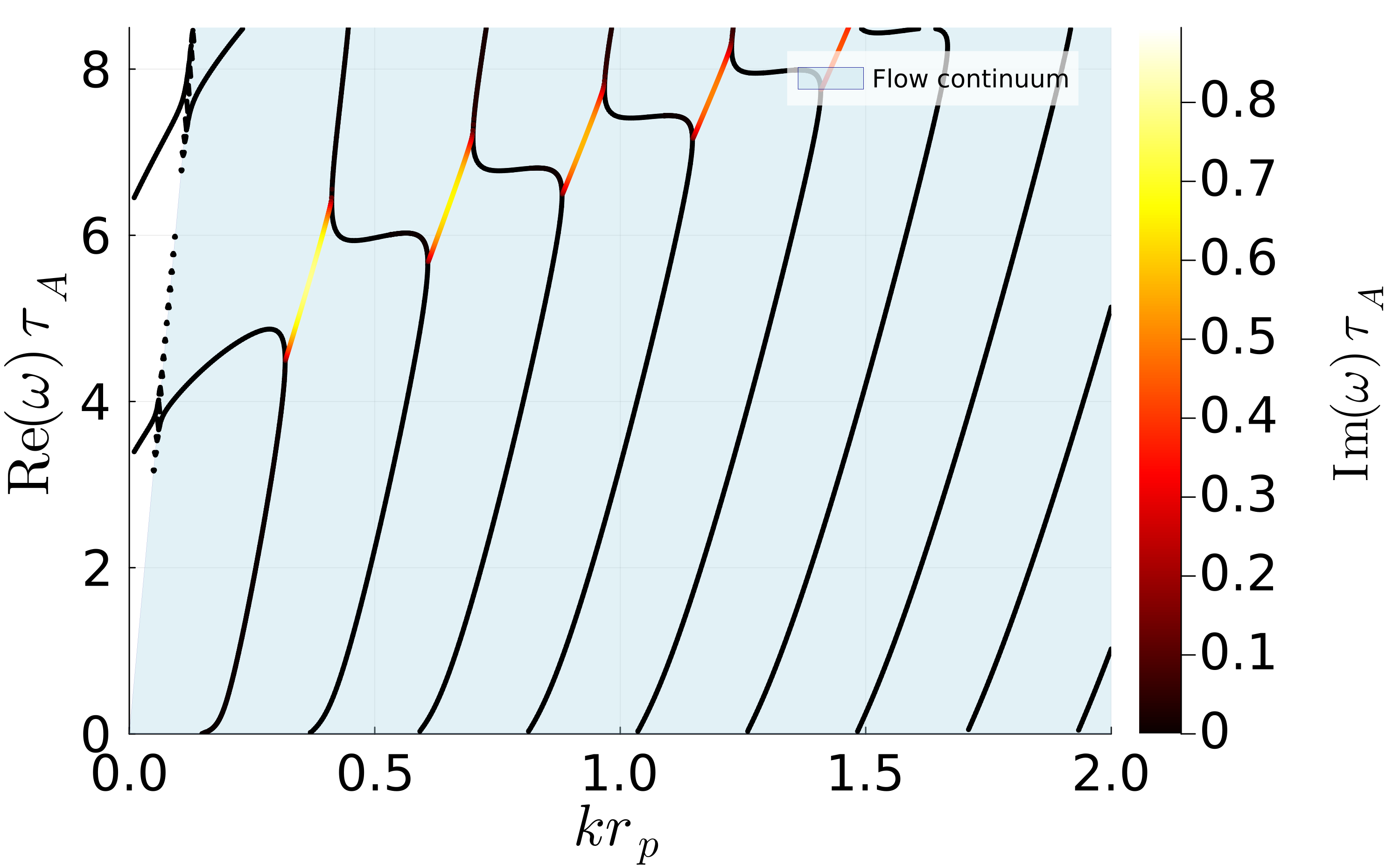}
    \caption{$M_A = 2$ ($v_0 = 4\,v_A^*$)}
    \label{fig:m0_v4}
\end{subfigure}
\caption{Dispersion relations at $m=0$ for Bennett equilibrium ($\alpha = 2$)
and wall at radius $r_w = 4\,r_p$.
The black curves are marginal modes computed by zero-contours of $\chi^+(\omega_r, k)$,
which shows some numerical artifacts at the edge of the continuous spectrum.
The flow continuum is shaded in light blue, which for $m=0$ arises from the Doppler shift $\omega = kv_{0z}(r)$.
The colored branches indicate unstable modes traced by continuation,
with color indicating growth rate $\text{Im}(\omega)$.}
\label{fig:m0_dispersion}
\end{figure*}

\begin{figure*}[htbp]
\centering
\begin{subfigure}[b]{0.48\textwidth}
    \includegraphics[width=\textwidth]{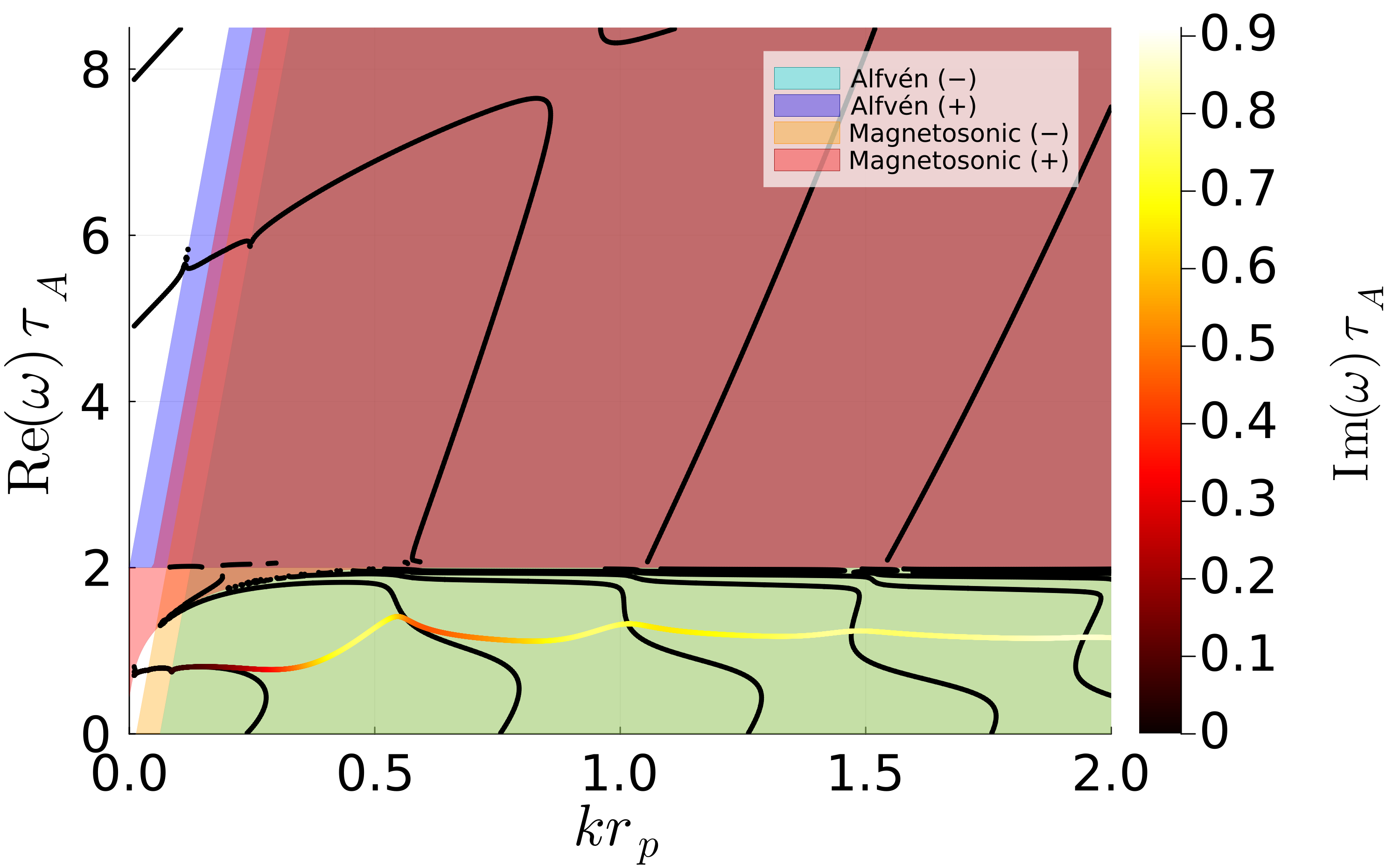}
    \caption{$M_A = 1$ ($v_0 = 2\,v_A^*$)}
    \label{fig:m1_v2}
\end{subfigure}
\hfill
\begin{subfigure}[b]{0.48\textwidth}
    \includegraphics[width=\textwidth]{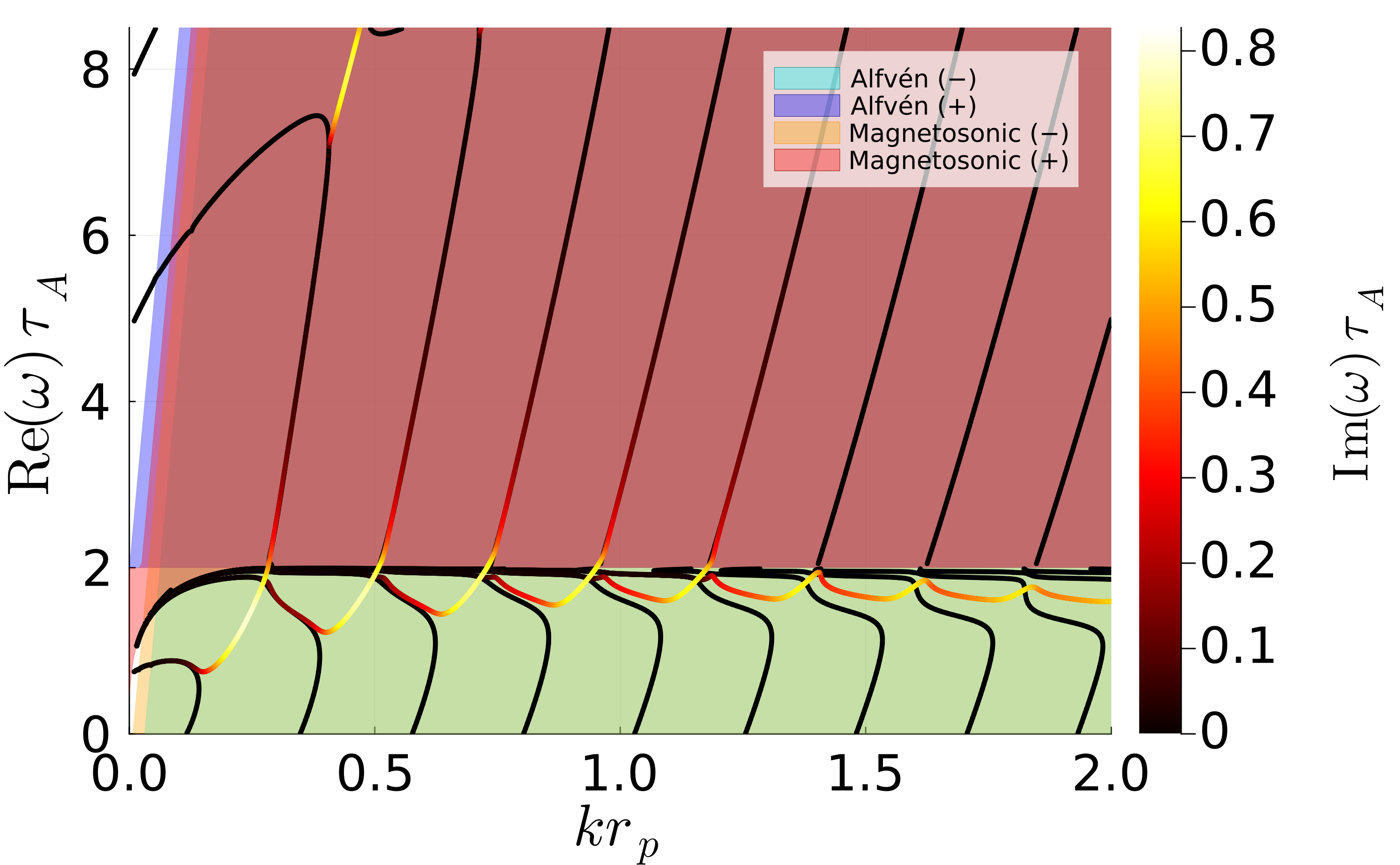}
    \caption{$M_A = 2$ ($v_0 = 4\,v_A^*$)}
    \label{fig:m1_v4}
\end{subfigure}
\caption{Computed dispersion relations with $m=1$ for Bennett equilibrium ($\alpha = 2$)
and wall at radius $r_w = 4\,r_p$.
Like Fig.~\ref{fig:m0_dispersion}, the black curves are zero-contours of $\chi^+(\omega_r, k)$,
the shaded regions are the four MHD continuum branches
(Alfv\'{e}n in teal/blue, magnetosonic in orange/red;
$\pm$ labels denote forward/backward Doppler-shifted branches; green is the overlap of teal and orange),
and the colored branches indicate unstable modes whose
color indicates the growth rate $\text{Im}(\omega)$.}
\label{fig:m1_dispersion}
\end{figure*}

\begin{figure}[htbp]
\centering
\includegraphics[width=\columnwidth]{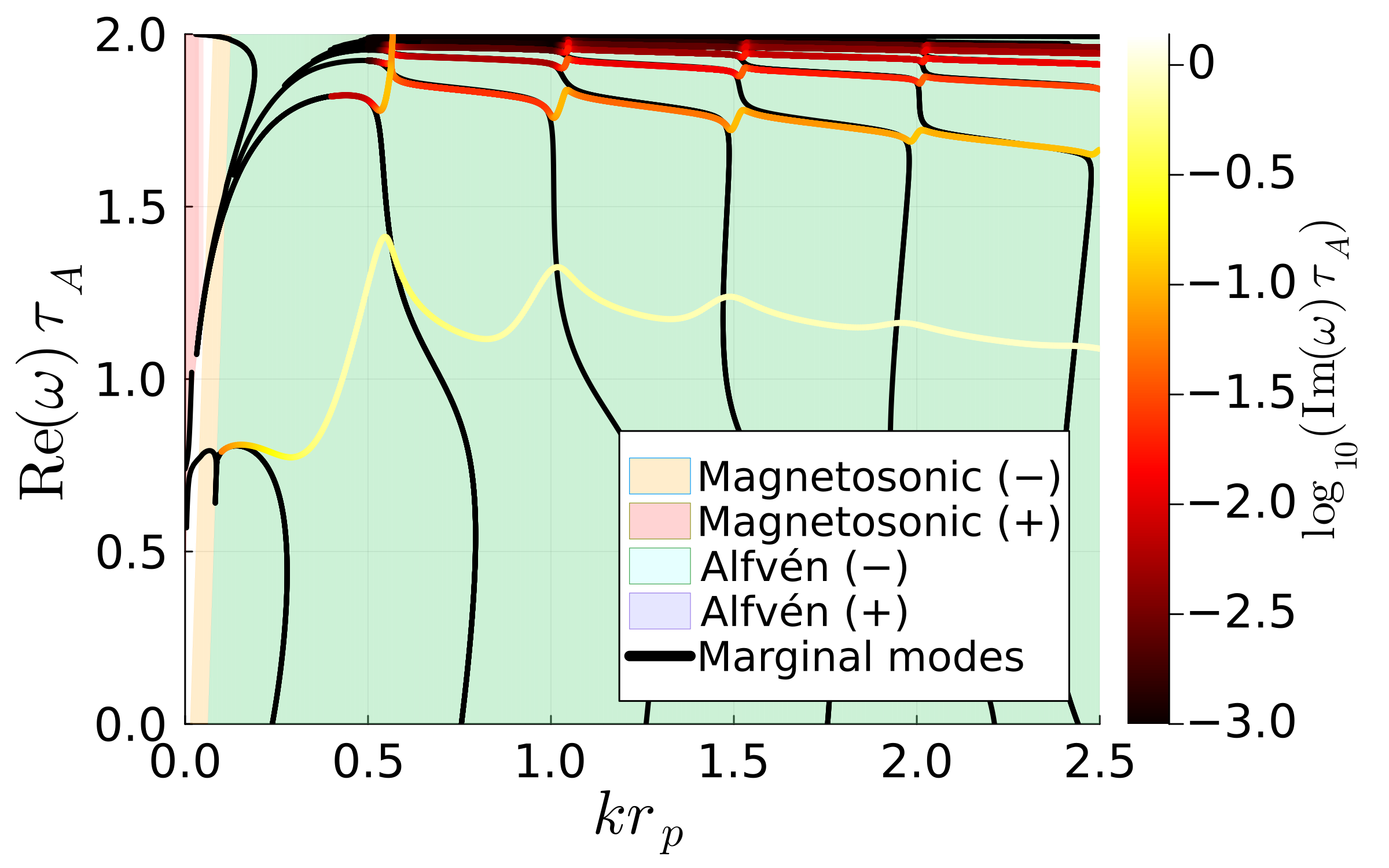}
\caption{Zoomed view of the $m=1$ dispersion relation at $M_A = 1$
(Fig.~\ref{fig:m1_v2}), showing the region near the Alfv\'{e}n frequency.
The modes accumulating near the Alfv\'{e}n frequency are
destabilized by coupling to exterior magnetosonic waves.
These tertiary instabilities have small growth rates
compared to the primary kink (here, with $\omega_r\tau_A\approx 1$) and reflection modes.}
\label{fig:m1_v2_zoom}
\end{figure}

The Krein-collision structure of Sec.~\ref{subsubsec:krein} also organizes the full
magnetohydrodynamic dispersion relations of Figs.~\ref{fig:m0_dispersion} and~\ref{fig:m1_dispersion}.
A Hamiltonian index theorem~\cite{Kapitula_2004} fixes the equilibrium's negative-energy content, so
sheared flow cannot stabilize the kink by removing it but must instead convert the unstable
eigenvalue into a negative-energy wave, which the same flow may later destabilize by collision.
In particular, the humps in the $M_A = 2$ curve of Fig.~\ref{fig:kink_growth_vs_k} are successive
Krein collisions of the negative-energy exterior branch with the positive-energy interior and axis
branches. The finite wall spaces these collisions closely enough that they overlap, so the growth
rate stays positive between the peaks and merges into a continuous unstable band as $r_w \to \infty$.

\section{Discussion and conclusion}
\label{sec:conclusion}

Within ideal MHD, sheared-flow Z-pinch stability is controlled 
by a competition between two spectral mechanisms, continuum damping and shear-driven excitation.
The interchange ($m=0$) and kink ($m=1$) modes are shear-stabilized by different continua.
For $m=0$, the Doppler-shifted continuum interacts
with the instability at all wavelengths, so sub-Alfv\'{e}nic sheared flow 
stabilizes even long wavelengths provided that the equilibrium 
is not too far from the Kadomtsev marginal profile.
For $m=1$, an ``Alfv\'{e}n spectral gap'' shields the long-wavelength kink
from continuum resonance, and trans-Alfv\'{e}nic flow is required
merely to close the gap and activate the damping mechanism.
This asymmetry explains why the $m=1$ mode is more stubborn to shear-flow stabilization 
than the $m=0$ mode, and is the central result of %
Secs.~\ref{subsec:MHD_continuum}--\ref{subsec:m0_stabilization}.

The trans-Alfv\'{e}nic sheared flow that exposes the kink to continuum damping simultaneously
excites shear-driven instabilities.
These modes include reflection modes from coupled acoustic cavities,
and an ``acoustic kink'' from coupling of the axis sound wave to external sound, 
as discussed in Sec.~\ref{subsec:shear_driven}.
But these are not MHD instabilities,
as they persist in the current-free limit and are essentially the same %
instabilities responsible for supersonic jet noise and turbulence.~\cite{Tam_1989,Tam_1995}
In MHD, the Alfv\'{e}n continuum provides additional
damping that may suppress weakly unstable reflection branches,
but the ``acoustic kink'' evades this damping
and persists across a broad range of wavenumbers.
These shear-driven modes emerge simultaneously with trans-Alfv\'{e}nic shear-flow stabilization
of the original MHD instabilities.

The shear-driven modes identified here are ideal, so
their fate under resistive or viscous dissipation and finite orbit-width effects
cannot be addressed by this analysis.
Higher fidelity simulations and more detailed examination are required.
Further, the present analysis is restricted to particular isothermal equilibria
with parabolic sheared-flow profiles;
general equilibria, including those with reversed shear
or non-monotone density, may admit additional phenomena.
In addition, it would be straightforward to extend 
the adiabatic/resonant splitting of the dispersion function
to the examination of sheared-flow screw pinches of various $q$-profiles.
Finally, the $|m| \geq 2$ flute modes, which possess even wider
shear-Alfv\'{e}n continuum gaps and correspondingly higher shear-stabilization thresholds,
may warrant additional study, although they are less deleterious to confinement
than the $m=1$ mode.
In summary, the spectral framework developed here provides a limiting foundation for an extended stability
theory of the sheared-flow Z~pinch including dissipation and kinetic effects,
and explains the persistent instability observed in ideal MHD eigensystem analysis.~\cite{Angus_2020}

\begin{acknowledgments}
  The authors would like to thank A.H. Glasser for suggesting the method to analyze
  the linearized MHD system as a function of a complex variable, U. Shumlak for many
  helpful discussions on Z-pinch stability and historical context, and E.T. Meier
  for clarifying the multifluid Z-pinch modeling literature.
\end{acknowledgments}

\section*{Author Declarations}

\subsection*{Conflict of Interest}
The authors have no conflicts to disclose.

\subsection*{Author Contributions}

\textbf{Daniel W. Crews}: Conceptualization (equal);
Formal analysis (equal);
Investigation (equal);
Methodology (equal);
Software (equal);
Validation (equal);
Visualization (equal);
Writing -- original draft (equal);
Writing -- review \& editing (equal).
\textbf{Jackson C. Turner}: Conceptualization (equal);
Formal analysis (equal);
Investigation (equal);
Methodology (equal);
Software (equal);
Validation (equal);
Visualization (equal);
Writing -- original draft (equal);
Writing -- review \& editing (equal).

\section*{Data Availability}
The data that support the findings of this study are available
from the corresponding author upon reasonable request.

\bibliography{references}

\appendix

\section{Regularization of the dispersion function}\label{app:regularization}
This appendix details regularization of the dispersion function $\chi(\omega, k)$ 
at the resonant surfaces of the Z-pinch equilibrium.
Although the resonances of $m \neq 0$ and $m=0$ present poles of different orders, 
the adiabatic and resonant dispersion functions are ultimately computed in the same way in both
cases, and indeed, due to a pole of any order.~\cite{Galapon_2016}

This appendix is structured as follows. 
Section~\ref{subsec:isolation} represents the integration operator (the propagator) 
as an ordered infinitesimal product to isolate the poles encountered during integration.
The point is that the integration through a resonant surface can be analyzed using 
standard techniques for singular integrals.
The cases of simple and double poles are treated separately in Sections~\ref{subsec:regular1}
and~\ref{subsec:regular2}, and Section~\ref{subsec:regularized_functions} then defines the
dispersion functions $\chi_{\text{adi}}$ and $\chi_\text{res}$.

\subsection{Isolation of singular integrals around resonant surfaces}\label{subsec:isolation}
The propagator, Eq.~\ref{eq:integral_operator}, may be written as the limit $N\to\infty$ of 
an ordered product of $N$ infinitesimal advances by $\Delta r$,
\begin{equation}\label{eq:ordered_product}
  \mathcal{L} = \lim_{N\to\infty}\prod_{n=N}^{1} (I + L(r_n^*)\Delta r) \equiv \lim_{N\to\infty}\prod_{n=N}^1\mathcal{L}_{n\Delta r}^{(n+1)\Delta r},
\end{equation}
where $[0, r_w]$ is discretized by $\Delta r$ and $r_n^*\in [n\Delta r, (n+1)\Delta r]$.
Choose an interval $[r_c - \epsilon, r_c + \epsilon]$ around a resonant layer at $r = r_c$
where $2\epsilon=\Delta r$.
Denoting $\vec{u}_\ell$ as the integration up to the interval's left side, the integration through the resonant layer is
\begin{equation}\label{eq:propagator_factorization}
  \mathcal{L}_{r_c-\epsilon}^{r_c+\epsilon}\vec{u}_\ell
   = \int_{r_c-\epsilon}^{r_c+\epsilon}\frac{\hat{L}(r)\vec{u}_\ell dr}{(\tilde{\omega}^2-m^2\omega_a^2)(\tilde{\omega}^2-m^2\omega_{ms}^2)}
\end{equation}
where $\hat{L}\equiv LD/\rho^2(\gamma p + B_0^2/\mu_0)$ (cf. Eq.~\ref{eq:denominator_form}) and $\tilde{\omega}\equiv\omega-kv_{0z}$.

\subsection{Regularization of simple poles ($m \neq 0$)}\label{subsec:regular1}
For $m \neq 0$, the linear resonances at $\omega = kv_{0z} \pm m\omega_a$ and 
$\omega = kv_{0z} \pm m\omega_{ms}$ appear as four simple poles in $\omega$-space.
Each resonance $\tilde{\omega} = m\omega_s$ (with $\omega_s \in \{\pm\omega_a, \pm\omega_{ms}\}$)
determines the radius $r_c$ of a resonant layer via
\begin{equation}\label{eq:critical_radius}
  \omega - kv_{0z}(r_c) = m\omega_s(r_c).
\end{equation}
Isolating the contribution to Eq.~\ref{eq:propagator_factorization} from a single resonance $r_c\in[0,r_w]$, 
the singular integral has the form
\begin{equation}\label{eq:singular_integral}
  \mathcal{L}_{r_c-\epsilon}^{r_c+\epsilon}\vec{u}_\ell
   = \int_{r_\ell}^{r_r} \frac{\vec{g}(r)}{\omega - \omega_s^*(r)} dr
\end{equation}
where $\omega_s^*(r) \equiv kv_{0z}(r) + m\omega_s(r)$, 
the vector-valued $\vec{g}(r)$ collects the regular factors,
and $r_\ell$, $r_r$ are the left and right sides of the infinitesimal interval.

For monotonic $\omega_s^*(r)$ with $d\omega_s^*/dr \neq 0$ at the resonance,
Eq.~\ref{eq:critical_radius} has a unique solution $r_c=r_c(\omega)$ and the integral can be
transformed to a standard Cauchy form.
Changing variables to the local frequency, $z = \omega_s^*(r)$, so that $dz = (d\omega_s^*/dr) dr$,
\begin{equation}\label{eq:singular_frequency_form}
  \int_{r_\ell}^{r_r} \frac{\vec{g}(r)}{\omega - \omega_s^*(r)} dr
  = \int_{\omega_s^*(r_\ell)}^{\omega_s^*(r_r)} \frac{\vec{g}(r(z))}{\omega - z}\frac{dz}{d\omega_s^*/dr}.
\end{equation}
This integral has a simple pole at $z = \omega$.  
Considering the limits $\text{Im}(\omega)\to \pm 0$, the results of which we denote by
$(\mathcal{L}_{r_c-\epsilon}^{r_c+\epsilon}\vec{u}_\ell)^\pm$,
the principal value is extracted with Plemelj's relation,
\begin{equation}\label{eq:plemelj}
  \frac{1}{\omega \pm 0i - z} = \text{P.V.}\frac{1}{\omega - z} \mp i\pi\delta(\omega - z).
\end{equation}
Returning to the radial variable gives
\begin{equation}\label{eq:plemelj_result}
  (\mathcal{L}_{r_c-\epsilon}^{r_c+\epsilon}\vec{u}_\ell)^\pm
   = \text{P.V.}\int_{r_\ell}^{r_r} \frac{\vec{g}(r)dr}{\omega - \omega_s^*(r)}
    \mp \frac{i\pi \vec{g}(r_c)}{d\omega_s^*/dr|_{r_c}}
\end{equation}
where the principal value is understood to apply to the integration over 
frequencies in the sense of Eq.~\ref{eq:singular_frequency_form}.
The quantity $d\omega_s^*/dr|_{r_c}^{-1}$ in the residue has a direct physical interpretation.
Strong shear narrows the radial width of resonance, weakening coupling between the continuum 
and the global mode.

When $d\omega_s^*/dr = 0$ at a continuum extremum, the change of variables
to Eq.~\ref{eq:singular_frequency_form} breaks down.

\subsection{Regularization of second-order poles ($m = 0$)}\label{subsec:regular2}
With $m = 0$, a critical radius $r_c$ is determined via
\begin{equation}\label{eq:critical_radius_m0}
  \omega = kv_{0z}(r_c).
\end{equation}
Isolating the integration through this resonance gives
\begin{equation}\label{eq:singular_integral_m0}
  \mathcal{L}_{r_c-\epsilon}^{r_c+\epsilon}\vec{u}_\ell
   = \int_{r_\ell}^{r_r} \frac{\vec{g}(r)}{(\omega - kv_{0z}(r))^2} dr
\end{equation}
where $\vec{g}(r)$ again collects regular factors.
The pole is second-order (since, with $m=0$, a factor of $\tilde{\omega}^2$ cancels between 
the matrix elements $C_1$, $C_2$, $C_3$, and the denominator $D$),
which necessitates a different regularization than the simple poles.

For monotonic $v_{0z}(r)$ with $dv_{0z}/dr \neq 0$ at the resonance,
changing variables to the flow velocity $v = v_{0z}(r)$ gives
\begin{equation}\label{eq:singular_velocity_form}
  \mathcal{L}_{r_c-\epsilon}^{r_c+\epsilon}\vec{u}_\ell
   = \int_{v_{0z}(r_\ell)}^{v_{0z}(r_r)} \frac{\vec{g}(r(v))}{k^2(\omega/k - v)^2}\frac{dv}{dv_{0z}/dr}
\end{equation}
revealing a second-order pole at $v = \omega/k$ where the flow matches the phase velocity.
Such poles are regularized by the Hadamard technique, which first applies the identity
\begin{equation}
  \mathcal{L}_{r_c-\epsilon}^{r_c+\epsilon}\vec{u}_\ell
   = -\frac{d}{d\zeta}\int_{v_{0z}(r_\ell)}^{v_{0z}(r_r)} \frac{\vec{g}(r(v))}{(\zeta - v)}\frac{dv}{dv_{0z}/dr}
\end{equation}
where $\zeta = \omega/k$. 
Inserting Eq.~\ref{eq:plemelj} for $\text{Im}(\omega)\to \pm 0$ obtains 
\begin{equation}\label{eq:velocity_form}
\begin{aligned}
(\mathcal{L}_{r_c-\epsilon}^{r_c+\epsilon}\vec{u}_\ell)^\pm
   &= -\frac{d}{d\zeta}\text{P.V.}\int_{v_{0z}(r_\ell)}^{v_{0z}(r_r)} \frac{\vec{g}(r(v))}{k^2(\zeta - v)}\frac{dv}{dv_{0z}/dr}\\
   &\mp i\pi \int_{L}^R\frac{d}{dv}\Big(\frac{\vec{g}(v)}{dv_{0z}/dr}\Big)\delta(\zeta-v)dv.
\end{aligned}
\end{equation}
The first term is the Hadamard finite part integral, defined as
\begin{equation}
  \mathcal{H}\int_a^b\frac{f(z)}{(z-\zeta)^2}dz \equiv \frac{d}{d\zeta}\text{P.V.}\int_a^b\frac{f(z)}{z-\zeta}dz
\end{equation}
provided that the Cauchy principal value integral exists.
Returning from Eq.~\ref{eq:velocity_form} to the radial variable gives the form
\begin{equation}\label{eq:hadamard_result}
  \begin{aligned}
(\mathcal{L}_{r_c-\epsilon}^{r_c+\epsilon}\vec{u}_\ell)^\pm
   = \mathcal{H}&\int_{r_\ell}^{r_r} \frac{\vec{g}(r)}{(\omega - kv_{0z}(r))^2} dr\\
  &\mp i\pi\Big(\frac{1}{dv_{0z}/dr}\frac{d}{dr}\Big(\frac{\vec{g}(r)}{dv_{0z}/dr}\Big)\Big)\Big|_{r=r_c}
  \end{aligned}
\end{equation}
where the Hadamard principal part notation $\mathcal{H}$ really refers to the integration over velocities.
The $m=0$ residue, or continuum coupling strength, depends more sensitively than the $m=1$ residue on both the 
magnitude of $dv_{0z}/dr$ and the critical points of the flow profile $v_{0z}$, \textit{i.e.}, where $dv_{0z}/dr=0$.

\subsection{Adiabatic and resonant dispersion functions}\label{subsec:regularized_functions}
Let each pole encountered in the integration through the plasma be treated by
the prescriptions of Sections~\ref{subsec:regular1} and~\ref{subsec:regular2}.
Write schematically the dispersion functions obtained by these prescriptions (Eqs.~\ref{eq:plemelj_result} and \ref{eq:hadamard_result})
as $\chi^\pm \equiv \chi_{P} \mp i \chi_R$ where $\chi_{P}$ denotes the principal part and
$\chi_R$ the sum of half-residues from all the poles.
Form the symmetric combinations
\begin{align}
  \chi^+ + \chi^- &=  2\chi_P, \\
  \chi^+ - \chi^- &= -2i\chi_R.
\end{align}
In the sum, the residues cancel to form $\chi_{\text{adi}} = \text{Re}[\chi^+] = \chi_P$,
and in the difference, the principal values cancel to form $\chi_{\text{res}} = \text{Im}[\chi^+] = -\chi_R$.
These are precisely the adiabatic and resonant dispersion functions defined in Section~\ref{subsubsec:regularization}.

\section{Incompressible stability of axisymmetric flow}\label{app:rayleigh}
This appendix reviews the hydrodynamic stability of incompressible flows,
which is first recapitulated and then detailed in the subsections.
A necessary condition for incompressible instability
of a planar shear flow $\vec{v} = u_0(y)\hat{x}$ is Rayleigh's inflection
point criterion: $d^2u_0/dy^2 = 0$ at some interior point.~\cite{Drazin_2004}
The criterion is modified in axisymmetric flow, as Rayleigh himself recognized.~\cite{Rayleigh_1896}
For axisymmetric perturbations ($m=0$), the Laplacian $\nabla^2$ is
replaced by the Grad-Shafranov operator $\Delta^*\psi \equiv r^2\nabla\cdot(r^{-2}\nabla\psi)$,
with the necessary condition for instability $\Delta^*u_0 = 0$ at an interior point.
The parabolic shear flow $v_z = v_0(r/r_p)^2$ satisfies $\Delta^*v_z=0$,
is therefore free of axisymmetric vortex-stretching, and is stable to $m=0$ incompressible instability.
For three-dimensional perturbations ($m\neq 0$), Ref.~\citenum{Batchelor_1962} showed that the
criterion generalizes.
Specifically, the quantity $Q = ru_0'/(m^2 + k^2r^2)$ must possess an extremum.
The parabolic shear flow satisfies $dQ/dr > 0$ everywhere and is therefore
stable to perturbations of every azimuthal mode number in the incompressible limit ($|v_z|\ll c_s$).

\subsection{Axisymmetric perturbations and vortex stretching}
Axisymmetric perturbations are treated most simply from Rayleigh's equation for the Stokes streamfunction.
Consider an axisymmetric radially sheared axial flow $\vec{v} = v_z(r)\hat{z}$ with
axisymmetric perturbations $\vec{v}_1(r, z, t)$.
Constraining the perturbation to be incompressible permits the 
Stokes streamfunction representation with components $v_{1r} = -r^{-1}\partial_z\psi$ and $v_{1z} = r^{-1}\partial_r\psi$.
The vorticity $\omega_\theta \equiv (\nabla\times\vec{v}_1)\cdot\hat{\theta}$ then relates to $\psi$ by
$\omega_\theta = r^{-1}\partial_{zz}\psi - \partial_r(r^{-1}\partial_r\psi)$.
The vorticity equation is
\begin{equation}\label{eq:vorticity_eqn}
  \partial_t\omega_\theta + v_z(r)\partial_z\omega_\theta = \partial_r(r^{-1}\partial_r v_z)\partial_z\psi
\end{equation}
where the vortex-stretching driving term is proportional to the gradients in $v_z(r)$.
Fourier transforming Eq.~\ref{eq:vorticity_eqn} as $(z, t) \to (k, \omega)$ by
$\psi(r, z, t) = \varphi(r) \exp(ik(z - \zeta t))$, where $\zeta = \omega/k$ is the phase velocity
yields Rayleigh's equation in the cylinder
\begin{equation}\label{eq:rayleigh_cylinder}
  (u_0 - \zeta)(\Delta^*\varphi - k^2\varphi) - (\Delta^*u_0)\varphi = 0
\end{equation}
where $\Delta^* \equiv r\frac{d}{dr}\left(\frac{1}{r}\frac{d}{dr}\right)$
is the axisymmetric Grad-Shafranov operator.
Equation~\ref{eq:rayleigh_cylinder} is identical to the Rayleigh equation in a planar shear flow
with the operator $\nabla^2$ replaced by $\Delta^*$.

To obtain an inflection point criterion, one takes the weighted inner product
of Eq.~\ref{eq:rayleigh_cylinder} with $\varphi^*/r$ from the axis to the wall
($r^{-1}$ is the weighting function of the Sturm-Liouville problem for the Grad-Shafranov operator).
This yields the weak form
\begin{equation}\label{eq:weak_form}
  \int_0^{r_w} \frac{dr}{r}\left(\left|\frac{d\varphi}{dr}\right|^2 + \big(k^2 + 
  \frac{\Delta^*u_0}{u_0 - \zeta}\big)|\varphi|^2\right) = 0
\end{equation}
after integrating by parts.
The imaginary part of Eq.~\ref{eq:weak_form} is
\begin{equation}\label{eq:imag_part_criterion}
  \int_0^{r_w} \frac{dr}{r}\frac{\Delta^*u_0\,|\varphi|^2}{|u_0 - \zeta|^2} = 0
\end{equation}
which can only be satisfied if $\Delta^*u_0$ changes sign.
Therefore, $\Delta^*u_0 = 0$ at some point is a necessary condition for incompressible Kelvin-Helmholtz 
instability in an axisymmetric flow.
At sufficiently large radius this reduces to the inflection point criterion, but generically
the condition is modified by curvature.

The kernel of the operator ($\Delta^*\mathcal{K} = 0$) consists of
constant and quadratic flows, $\mathcal{K} = c_1r^2 + c_2$, while its eigenfunctions 
($\Delta^*\mathcal{E} = \lambda^2\mathcal{E}$) are the first-order Bessel functions, $\mathcal{E} = cJ_1(\lambda r)$.
The parabolic profile $u_0 \sim r^2$ is in the kernel of the operator, so that Rayleigh's equation
$(u_0-\zeta)(\Delta^* - k^2)\varphi = 0$ is homogeneous for such a profile.
This homogeneity means the incompressible eigenfunctions are particularly simple for such a profile,
lacking any driving term.
Physically, there is no incompressible vortex-stretching by such a flow,
but incompressible perturbations still exist as neutrally stable modes
advected by the shear.

\subsection{Three-dimensional perturbations and stability criterion}
For general perturbations of azimuthal mode number $m$ and axial
wavenumber $k$, the incompressible eigenvalue problem is governed by the
cylindrical Rayleigh equation of Ref.~\citenum{Batchelor_1962},
\begin{equation}\label{eq:bg_rayleigh}
  (u_0 - \zeta)\Big(\frac{d}{dr}\!\left(\frac{r}{\mu^2}\frac{d(rG)}{dr}\right)
  - \,G\Big)
  - rG\,\frac{dQ}{dr} = 0,
\end{equation}
where $G(r)$ is the radial velocity,
$\mu^2(r) \equiv m^2 + k^2r^2$ is the squared transverse wavenumber,
and $Q(r) = ru_0'/\mu^2$ collects a weighted vorticity gradient.
Reference~\citenum{Batchelor_1962} shows that the necessary condition for instability 
obtained from Eq.~\ref{eq:bg_rayleigh} is that the quantity $Q$ must have an
extremum, $dQ/dr = 0$.

Differentiating $Q$ yields
\begin{equation}\label{eq:Q_decomposition}
  \frac{dQ}{dr}
  = \frac{r\,\Delta^* u_0}{m^2 + k^2r^2}
  + \frac{2m^2\, u_0'}{(m^2 + k^2r^2)^2}\,,
\end{equation}
separating into a vortex-stretching term proportional to $\Delta^*u_0$
and a curvature-coupling term proportional to $m^2$.
For $m = 0$ the second term vanishes and the criterion
reduces to $\Delta^*u_0 = 0$, recovering the result of the previous
subsection.
For $m \neq 0$ the curvature coupling introduces an additional contribution
that is positive-definite when $u_0' > 0$; consequently, a monotonically
increasing flow in which vortex stretching vanishes
($\Delta^*u_0 = 0$, such as $u_0 \sim r^2$) satisfies
$dQ/dr > 0$ everywhere and is stable to incompressible perturbations of
\emph{every} azimuthal mode number.

The criterion of Ref.~\citenum{Batchelor_1962} acquires a geometric interpretation 
by introducing a helical eikonal coordinate
\begin{equation}\label{eq:helical_eikonal}
  d\eta = \frac{m^2 + k^2r^2}{r}\,dr
  = k_\perp^2\,r\,dr,
  \qquad
  \eta = m^2\ln r + \tfrac{1}{2}k^2r^2,
\end{equation}
where $k_\perp(r) = \sqrt{m^2/r^2 + k^2}$ is the local transverse
wavenumber on the helices of constant phase discussed in
Ref.~\citenum{Batchelor_1962}.
In this coordinate one verifies that $Q = du_0/d\eta$, so the necessary
condition $dQ/dr = 0$ becomes simply
$d^2u_0/d\eta^2 = 0$, analogous to the original Rayleigh criterion.
The physical meaning is that the velocity profile must possess an inflection point 
when measured in accumulated eikonal phase.
For $m = 0$ the eikonal reduces to $\eta \propto r^2$ and the condition
recovers $\Delta^*u_0 = 0$; for $m \neq 0$ the logarithmic piece
$m^2\ln r$ warps the coordinate near the axis, encoding a centrifugal
barrier that expels helical perturbations from the core.
Note that no axis-regular flow profile can satisfy $dQ/dr=0$ for $m\neq 0$.

\section{Acoustic modes in compressible shear flow}\label{app:wkb}
Compressible instabilities arise from the resonant coupling of sound waves
in high Mach number shear flows.
This appendix discusses the acoustics of compressible perturbations
of a current-free axisymmetric shear flow, providing an analytical framework for the
reflection mode instabilities discussed in Section~\ref{subsec:shear_driven}.
It should be understood that the following strictly applies to jets in the 
high Reynolds number limit $\text{Re}\to\infty$.

\subsection{Reduced system for the current-free Z pinch}
Specialize the general operator $L$ (Eq.~\ref{eq:linear_operator}) to the current-free
equilibrium $B_{0\theta}=B_{0z}=0$ with uniform density and pressure ($\rho_0 = p_0 = 1$)
and general azimuthal mode number $m$.
Then $F = \vec{k}\cdot\vec{B}_0 = 0$, $C_1 = 0$, and the first-order system
$d\vec{u}/dr = L\vec{u}$ with $\vec{u} = (r\xi_r, P)$ reduces to
\begin{align}
  \partial_r(r\xi_r) &= -r\left(\frac{1}{c_s^2}
    - \frac{k^2 + m^2/r^2}{\tilde{\omega}^2}\right) P, \label{eq:wkb_sys1}\\
  \partial_r P &= \tilde{\omega}^2\,\xi_r, \label{eq:wkb_sys2}
\end{align}
where $\tilde{\omega}(r) = \omega - k v_{0z}(r)$ is the Doppler-shifted frequency
and $c_s^2 = 2\gamma$ in the normalized (Alfv\'{e}nic) units, so $c_s \approx 1.83\,v_A^*$.
This is the compressible hydrodynamic system for a current-free
radially sheared axial flow, describing acoustic propagation
and, in the limit of vanishing Mach number, the incompressible vortical dynamics.

\subsection{Second-order equation and WKB dispersion relation}
Eliminating $r\xi_r$ by substituting Eq.~\ref{eq:wkb_sys2} into Eq.~\ref{eq:wkb_sys1}
yields a second-order equation for the total pressure perturbation,
\begin{equation}\label{eq:pridmore_brown}
  \frac{\tilde{\omega}^2}{r}\frac{d}{dr}\Big(\frac{r}{\tilde{\omega}^2}\frac{dP}{dr}\Big) + \Big(\frac{\tilde{\omega}^2}{c_s^2} - \big(k^2 + \frac{m^2}{r^2}\big)\Big)P=0
\end{equation}
known as the Pridmore-Brown equation.~\cite{Pridmore_Brown_1958,Rienstra_2020}

Expanding the self-adjoint operator in Eq.~\ref{eq:pridmore_brown} produces
a second-order ODE with a first-derivative term proportional
to $v_{0z}'/\tilde{\omega}$ (the very term whose singularity at $\tilde{\omega}=0$
generates the resonances of Sec.~\ref{sec:marginal}).
In the WKB limit of short radial wavelength, and away from a resonance, 
this shear-coupling term can be neglected.
Then seeking solutions $P \sim \exp(i\int k_r\,dr)$ produces the local dispersion relation
\begin{equation}\label{eq:acoustic_dispersion}
  \tilde{\omega}^2 = c_s^2(k_r^2 + k_\theta^2 + k^2)
\end{equation}
where $k_\theta = m/r$ is the azimuthal wavevector.

Equation~\ref{eq:acoustic_dispersion} is inaccurate near the axis where
the geometric term $r^{-1}dP/dr$ is non-negligible.
Substituting $P = r^{-1/2}\,u$ removes this singularity
and reduces Eq.~\ref{eq:pridmore_brown} to the standard WKB form
$u'' + k_r^2\,u = 0$
(with the shear-coupling term dropped), 
resulting in a corrected dispersion relation
\begin{equation}\label{eq:langer}
  k_r^2(r) = \frac{\tilde{\omega}^2}{c_s^2}
    - k^2 - \frac{m^2 - \tfrac{1}{4}}{r^2}.
\end{equation}
The shift $m^2 \to m^2 - \tfrac{1}{4}$ is known as the Langer correction
and better approximates the WKB phase with the exact Bessel-function
behavior near $r = 0$.~\cite{Langer_1937}
The Langer correction is only used for $|m| \geq 1$,
since the axis is a regular point for $m=0$ and the substitution is invalid there.

\subsection{Mode trapping and quantization}
This section derives the quantization conditions underlying the three Families of Sec.~\ref{sec:three_family},
where Family (i) are interior acoustic modes propagating between the axis 
(or the ``centrifugal'' barrier for $|m|\geq 1$) and a turning point, 
Family (ii) are exterior acoustic modes propagating between the sonic point and the wall,
and Family (iii) for $|m|\geq 1$ is an axis-localized mode.
The turning point occurs where $k_r^2 = 0$ in Eq.~\ref{eq:acoustic_dispersion} for $m=0$
or in Eq.~\ref{eq:langer} for $|m|\geq 1$.
Modes propagate where $k_r^2>0$ and are evanescent where $k_r^2<0$.

\paragraph{Interior family.}
The interior modes are forward-propagating sound waves
trapped in the subsonic core, turned back at $R_1$ where the
co-moving speed drops to the local sound speed 
($\tilde{\omega} = +c_s k$ for $m=0$).
The sonic point $R_1$ is defined by
\begin{equation}\label{eq:turning_point}
  \frac{\tilde{\omega}(R_1)^2}{c_s^2} = k^2 + \frac{m^2 - \tfrac{1}{4}}{R_1^2}
\end{equation}
without the factor $1/4$ for $m = 0$.
For $m = 0$ the interior region spans from the axis to $R_1$,
and the quantization condition is
\begin{equation}\label{eq:quantization}
  \int_0^{R_1} k_r(r)\,dr = \left(n + \tfrac{1}{2}\right)\pi, \qquad n = 0, 1, 2, \ldots,
\end{equation}
where the $1/2$ accounts for the $\pi/4$ phase shifts at the axis and turning point.~\cite{Bender_Orszag_1999}
For $m \neq 0$, the divergent azimuthal wavevector $k_\theta = m/r$
drives $k_r^2$ negative near the axis, producing an inner turning point $r_{\min}$ 
satisfying Eq.~\ref{eq:turning_point}.
The wave propagates only in the annulus $r_{\min} < r < R_1$,
giving
\begin{equation}\label{eq:quantization_m}
  \int_{r_{\min}}^{R_1}
    \sqrt{\frac{\tilde{\omega}^2}{c_s^2} - k^2
      - \frac{m^2 - \tfrac{1}{4}}{r^2}}\,dr
  = \left(n + \tfrac{1}{2}\right)\pi.
\end{equation}
with $n = 0, 1, 2, \ldots$.
Each value of $n$ selects a discrete frequency $\omega_n(k)$,
producing the interior branches in the current-free dispersion
relation shown in Figs.~\ref{fig:current_free_m0}b and~\ref{fig:current_free_m1}b.
At $k = 0$ the equation reduces to Bessel's equation
(Sec.~\ref{subsec:surface_mode}) and the eigenfrequencies are
known exactly: $\omega_n = c_s\, j_{1,n}/r_w$ for $m = 0$
and $\omega_n = c_s\, j'_{1,n}/r_w$ for $m = 1$,
where $j_{1,n}$ and $j'_{1,n}$ are zeros of $J_1$ and $J_1'$ respectively.
These exact values calibrate the WKB phase integrals
in the figures, improving accuracy for the lowest-order branches
where the asymptotic index $(n + \tfrac{1}{2})\pi$ incurs $O(1/n)$ errors.

\paragraph{Exterior family.}
The exterior modes are backward-propagating sound in the
fast-flow exterior which have been Doppler-shifted by the locally supersonic flow 
to be forward-propagating in the lab frame.
These modes reflect back outward at a turning point $R_2$
where $k_r^2 = 0$.
The exterior cavity extends from $R_2$ to the wall at $r_w$,
with quantization condition
\begin{equation}\label{eq:exterior_quantization}
  \int_{R_2}^{r_w}
    \sqrt{\frac{\tilde{\omega}^2}{c_s^2} - k^2
      - \frac{m^2 - \tfrac{1}{4}}{r^2}}\,dr
  = \left(j + \tfrac{1}{2}\right)\pi,
\end{equation}
for $j = 0, 1, 2, \dots$.
These branches appear as steep, nearly vertical curves in the $(k,\omega)$ plane,
rising through $\omega = 0$ from negative frequencies at small $k$
seen in Figs.~\ref{fig:current_free_m0}b and~\ref{fig:current_free_m1}b.
The discrete modes merge into a radiation continuum as $r_w\to\infty$.

\subsection{Reflection instability and jet screech}\label{subsec:two_cavity}
Because the exterior family is Doppler-shifted by the shear flow to
propagate forwards, the interior and exterior families may come into resonance, 
as discussed in Sec.~\ref{sec:marginal}.
The sheared flow separates the modes by an evanescent barrier from $R_1$ to $R_2$, 
which contains the critical layer $\tilde{\omega} = 0$.
Such instabilities are called reflection modes.

The instability occurs where the interior and exterior branches
intersect in the $(k,\omega)$ plane.
At such a crossing, the two cavities share the same frequency and wavenumber,
coupling resonantly through the evanescent barrier.
Each intersection of an interior branch $n$ with an exterior branch $j$
produces instability in the numerical dispersion relation,
explaining the pattern of ``humps'' along each dispersion branch
in Figs.~\ref{fig:current_free_m0}a and~\ref{fig:current_free_m1}a.
As $r_w\to\infty$, the exterior family merges into a continuum
and the discrete humps give way to a continuous band of instability.

The reflection modes are the local instability mechanism underlying
``jet screech,'' in which supersonic jets radiate intense
sound at discrete frequencies selected by downstream feedback.~\cite{Tam_1995}
Squire's theorem does not hold in an axisymmetric geometry~\cite{Batchelor_1962}
and the dominant reflection mode is typically not $m=0$
in high-Reynolds number supersonic jets.

\subsection{Axis mode (Family~iii) and acoustic kink}\label{subsec:surface_mode}
Any regular axial flow satisfies $v_z = f(r^2)$,
so $v_{0z}'(0) = 0$ and the shear-coupling term
in the Pridmore-Brown equation vanishes at $O(r)$ near the axis.
For $|m|\geq 1$ the near-axis equation is therefore
Bessel's equation
\begin{equation}\label{eq:bessel_axis}
  P'' + \frac{1}{r}P' + \left(k_{r0}^2 - \frac{m^2}{r^2}\right)P = 0,
  \qquad k_{r0}^2 \equiv \frac{\omega^2}{c_s^2} - k^2,
\end{equation}
whose regular solution is $P \propto J_m(k_{r0}\, r)$ for $\omega > k c_s$
and radially quantizes.
When the axial phase velocity is sub-sonic ($\omega/k < c_s$),
$k_{r0}^2 < 0$ and $P\propto I_m(k_{r0}\, r)$,
meaning the mode is not radially quantized as it propagates only azimuthally and axially.
This axis mode is destabilized by resonantly coupling with the exterior acoustic modes, Family (ii).
In the case $m=1$ this manifests as an acoustic kink instability, 
arising even for profiles satisfying the incompressible stability condition of Appendix~\ref{app:rayleigh}.

\end{document}